\documentstyle[pre,aps]{revtex}

\squeezetable

\begin{document}

\draft

\author{V. I. Yukalov$^{1,2}$ and S. Gluzman$^3$}
\address{$^1$Centre for Interdisciplinary Studies in Chemical Physics \\
University of Western Ontario, London, Ontario N6A 3K7, Canada \\
$^2$Bogolubov Laboratory of Theoretical Physics \\
Joint Institute for Nuclear Research, Dubna 141980, Russia \\
$^3$International Centre of Condensed Matter Physics\\
University of Brasilia, CP 04513, Brasilia, DF 70919--970, Brazil}

\title{Self--Similar Exponential Approximants}

\maketitle

\begin{abstract}
An approach is suggested defining effective sums of divergent series in the
form of self--similar exponential approximants. The procedure of constructing
these approximants from divergent series with arbitrary noninteger powers is
developed. The basis of this construction is the self--similar approximation
theory. Control functions governing the convergence of exponentially
renormalized series are defined from stability and fixed--point conditions
and from additional asymptotic conditions when the latter are available. The
stability of the calculational procedure is checked by analyzing cascade
multipliers. A number of physical examples for different statistical systems
illustrate the generality and high accuracy of the approach.
\end{abstract}

\vspace{1cm}


\section{ Introduction}

Summation of divergent series is the problem of great importance in
theoretical physics, applied mathematics and engineering. This is because
realistic problems are usually solved by means of some calculational 
algorithm often resulting in divergent sequences of approximations. 
Assigning a finite value to the limit of a divergent sequence is called
renormalization or summation technique. The most widely used such technique
is Pad\'e summation [1]. However, the latter has several shortcomings.
First of all, to reach a reasonable accuracy of Pad\'e
approximants, one needs to know at least in the magnitude of ten to twenty
terms of a perturbation series. Unfortunately, so many terms are often not
available because of the complexity of a considered problem. Second, Pad\'e
approximants are defined for the series of integer powers. But in many cases
asymptotic series arise having noninteger powers. Third, there are quite
simple examples [2] that are not Pad\'e summable even for a sufficiently
small variable. Moreover, the standard Pad\'e approximants do not converge
at infinity, since infinity is an essential singularity [1]. The latter
deficiency can be sometimes treated with the help of two--point Pad\'e
approximants [3]. But these can only be constructed when there are two
perturbation expansions in the vicinity of two points, so that such
expansions have compatible variables, which often is not the case,
especially when one of or both these expansions contain noninteger powers
[4]. For instance, only rational powers can be described at infinity [3].
This is because a Pad\'e approximant is a ratio of two polynomials, say,
of $m$ and $n$ order. Therefore, when the variable tends to infinity, the
asymptotic behaviour of this Pad\'e approximant is of power law with the
power $m/n$, which is a rational number. Fifth, in many physical problems
the quantities of interest exhibit at infinity exponential behaviour,
which in principle cannot be described by Pad\'e approximants. One more
well known difficulty when dealing with Pad\'e approximants is the 
appearance of spurious poles [1]. The last, but not the least, Pad\'e
summation is rather a numerical technique providing answers in the form of
numbers. Therefore, it is difficult, if possible, to analyze the results
when the considered problem contains several parameters to be varied, since
for each given set of parameters one has to repeat the whole procedure of
constructing a table of Pad\'e approximants and selecting of
them one corresponding to a visible saturation of numerical values. Being a
numerical technique, Pad\'e summation shares the difficulties
of other numerical methods, like nonlinear sequence transformations, and
sometimes is less effective than the latter [5,6].

The aim of the present paper is to develop an{\it \ analytical} method, 
free of the Pad\'e approximation difficulties, for
summing divergent series containing any number of terms ( just a few or
many) with arbitrary {\it noninteger} powers. The method is based on the
ideas of the self--similar approximation theory [7-14] in its algebraically
invariant formulation [12-14]. The novelty of this paper is in what follows:

(i)  The approach is generalized by constructing self--similar 
exponential approximants from the series with {\it arbitrary powers}, 
integer as well as noninteger. (ii) It is shown how such exponential 
approximants can be made compatible with additional {\it asymptotic} and
{\it boundary conditions}. (iii) Stability and fixed--point conditions 
are discussed and concrete prescriptions, for defining control functions 
and for checking the {\it stability} of the calculational procedure, are 
formulated. (iv) It is demonstrated that the approach is applicable in 
all cases, when either just a few terms of a series are known or when 
{\it many terms} are available. (v) We emphasize that in all the cases, 
even when a large number of terms of a series are involved, we obtain 
{\it analytical formulae} that are convenient for considering with respect
to a change of physical parameters. The possibility of deriving analytical
expressions is a characteristic feature of our approach differing it from 
numerical methods.

Also, we illustrated the generality of the approach by applying it to
several interesting physical problems of quite different nature. Some of 
these problems could be treated by Pad\'e summation or by the standard 
renormalization--group technique. When such treatments have been done, we 
compare their results with ours. In those cases for which numerical data 
are available, we estimate the accuracy and show that this accuracy is either
comparable or higher than that of other more complicated resummation 
techniques. We construct as well self--similar exponential approximants 
for several physical problems for which other summation methods have not 
been applied because of technical difficulties.

\section{Noninteger Powers}

The main idea [7] of the self--similar approximation theory is to put into
correspondence to a sequence of perturbative terms a dynamical system for
which the approximation number would play the role of discrete time. Then
the transformation from one approximation to another can be represented as
the evolution of this dynamical system [7]. The corresponding evolution
equation may be formulated as the property of functional self--similarity,
which is a necessary condition of fast convergence [8]. A dynamical system
with discrete time is called a cascade. A fixed point of a cascade,
representing a sequence of approximations, corresponds to an effective
limit of this sequence [7,8]. To guarantee convergence, the fixed point
must be stable and the sufficient conditions of stability can be
formulated in terms of multipliers [9-11]. Additional possibilities open
if we require that the procedure of finding an effective limit of a
sequence is invariant with respect to algebraic transformations [12-14],
which permits us to deal not with the initial sequence of approximations
but with a sequence of its transforms. General ideas and the mathematical
foundation of the self--similar approximation theory have been described
in detail in our previous papers [7-14]. Not repeating them here, we begin
at once with considering the case when for a sought function $f(x)$ one
derives an approximate perturbative expression 
\begin{equation}
\label{1}p_k(x)=\sum_{n=0}^ka_n\ x^{\alpha _n}, 
\end{equation}
in which $\alpha _n$ are arbitrary real numbers, integer or noninteger,
positive or negative, but with the sole requirement that they form an
ordered sequence $\{\alpha_n\}$, that is, either strictly increasing or
strictly decreasing sequence of terms. Following the method of the
algebraic self--similar renormalization [12-14], we define the algebraic
transform 
\begin{equation}
\label{2}P_k(x,s)\equiv x^s\ p_k(x)=\sum_{n=0}^ka_n\ x^{s+\alpha _n}, 
\end{equation}
where $s$ is real. Then, by means of the equation 
\begin{equation}
\label{3}P_0(x,s)=a_0\ x^{s+\alpha _0}=\varphi , 
\end{equation}
we obtain the expansion function 
\begin{equation}
\label{4}x(\varphi ,s)=\left( \frac \varphi {a_0}\right) ^{1/\left( s+\alpha
_0\right) }. 
\end{equation}
Substituting the latter into Eq. (2), we have 
\begin{equation}
\label{5}y_k(\varphi ,s)\equiv P_k(x(\varphi ,s),s)=\sum_{n=0}^ka_n\ \left(
\frac \varphi {a_0}\right) ^{\left( s+\alpha _n\right) /\left( s+\alpha
_0\right) }. 
\end{equation}
The family $\{y_k\}$ of transforms (5) is called [10, 11] the approximation
cascade, since its trajectory $\{y_k(\varphi ,s)\mid k=0,1,2...\}$ is
bijective to the sequence $\{P_k(x,s)\mid k=0,1,2...\}$ of approximations
(2). A cascade is a dynamical system in discrete time $k=0,1,2...,$ whose
trajectory points satisfy the semigroup property $y_{k+p}(\varphi
,s)=y_k(y_p(\varphi ,s),s).$ The latter equation is a particular type of
functional equations [15] and in dynamical theory it is related to
autonomous dynamical systems [16]. The physical meaning of the above
semigroup relation can be understood as the property of functional
self--similarity [17] with respect to the varying approximation number
[7-11]. The self--similarity relation is a necessary condition for the
fastest convergence criterion [8,9].

For the approximation cascade $\{y_k\},$ defined by transform (5), the
cascade velocity can be written as a finite difference
\begin{equation}
\label{6}v_k(\varphi ,s)\equiv y_k(\varphi ,s)-y_{k-1}(\varphi ,s)=a_k\
\left( \frac \varphi {a_0}\right) ^{\left( s+\alpha _k\right) /\left(
s+\alpha _0\right) }. 
\end{equation}
This is to be substituted into the evolution integral 
\begin{equation}
\label{7}\int_{P_{k-1}}^{P_k^{*}}\frac{d\varphi }{v_k(\varphi ,s)}=\tau , 
\end{equation}
in which $P_k=P_k(x,s)$ and $\tau $ is the minimal time needed for reaching
a fixed point $P_k^{*}=P_k^{*}(x,s,\tau ).$ Integral (7) with velocity (6)
yields 
\begin{equation}
\label{8}P_k^{*}(x,s,\tau )=\left[ P_{k-1}^{-\nu }(x,s)-\frac{\nu \ a_k\tau 
}{a_0^{1+\nu }}\right] ^{-1/\nu }, 
\end{equation}
where 
\begin{equation}
\label{9}\nu =\nu _k(s)\equiv \frac{\alpha _k-\alpha _0}{s+\alpha _0}. 
\end{equation}
Taking the algebraic transform, inverse to Eq.(2), we find 
\begin{equation}
\label{10}p_k^{*}(x,s,\tau )\equiv x^{-s}P_k^{*}(x,s,\tau )=\left[
p_{k-1}^{-\nu }(x)-\frac{\nu \ a_k\tau }{a_0^{1+\nu }}x^{s\nu }\right]
^{-1/\nu }. 
\end{equation}
Exponential renormalization [13,14] corresponds to the limit $s\rightarrow
\infty$, at which 
$$\lim _{s\rightarrow \infty }\nu _k(s)=0 , \qquad 
\lim_{s\rightarrow \infty }s\nu _k(s)=\alpha _k-\alpha _0 . $$ 
Then Eq. (10) gives \begin{equation}
\label{11}\lim _{s\rightarrow \infty }p_k^{*}(x,s,\tau )=p_{k-1}(x)\exp
\left( \frac{a_k}{a_0}\tau x^{\alpha _k-\alpha _0}\right) . 
\end{equation}
Accomplishing exponential renormalization of all sums appearing in
expressions of type (11), we follow the bootstrap procedure [14] according to
the scheme 
\begin{equation}
\label{12}p_k(x)\rightarrow p_k^{*}(x,s,\tau )\rightarrow F_k(x,\tau _1,\tau
_2,...,\tau _k), 
\end{equation}
with $k\geq 1.$

Let us illustrate explicitly how this exponential bootstrap works. The
initial approximation from Eq.(1) is
\begin{equation}
\label{13}p_0(x)=a_0x^{\alpha _0}. 
\end{equation}
If we limit ourselves by the first-order term%
$$
p_1(x)=p_0(x)+a_1x^{\alpha _1}, 
$$
then, following the renormalization scheme (12), we get 
\begin{equation}
\label{14}F_1(x,\tau _1)=p_0(x)\exp \left( b_1x^{\beta _1}\right) , 
\end{equation}
where 
\begin{equation}
\label{15}b_1\equiv \frac{a_1}{a_0}\tau _1,\quad \beta _1\equiv \alpha
_1-\alpha _0\ . 
\end{equation}
Involving the second order term%
$$
p_2(x)=p_1(x)+a_2x^{\alpha _2}, 
$$
we find 
\begin{equation}
\label{16}F_2(x,\tau _1,\tau _2)=p_0(x)\exp \left( b_1x^{\beta _1}\exp
\left( b_2x^{\beta _2}\right) \right) , 
\end{equation}
with the notation 
\begin{equation}
\label{17}b_2\equiv \frac{a_2}{a_1}\tau _2,\quad \beta _2\equiv \alpha
_2-\alpha _1. 
\end{equation}
Continuing the same procedure, for the $k$--th order expression (1) we
obtain 
\begin{equation}
\label{18}F_k(x,\tau _1,\tau _2,...,\tau _k)=p_0(x)\exp \left( b_1x^{\beta
_1}\exp \left( b_2x^{\beta _2}...\exp \left( b_kx^{\beta _k}\right) \right)
...\right) , 
\end{equation}
where 
\begin{equation}
\label{19}b_k\equiv \frac{a_k}{a_{k-1}}\tau _k,\quad \beta _k\equiv \alpha
_k-\alpha _{k-1}. 
\end{equation}

The quantities $\tau _n,$ with $n=1,2,\ldots k$, in the renormalized form (18)
play the role of control functions [7-14]. In the following section we shall
show the ways of defining these control functions. Assume, for a while, that
the latter have already been defined giving $\tau _n=\tau _n(x).$
Substituting these functions into Eq.(18), we come to the self-similar
exponential approximant 
\begin{equation}
\label{20}f_k^{*}(x)=F_k(x,\tau _1(x),\tau _2(x),...,\tau _k(x)). 
\end{equation}
Note that the constructed approximants (20) are different from the iterated 
exponentials introduced by Euler [18]; the properties of such iterated
exponentials being reviewed, e.g., in Refs. [19,20].

Although in this paper we shall deal with physical problems
related to series with powers being real numbers, nothing prevents us from
generalizing the whole approach to series with complex powers. Such complex
exponents appear in the problems with discrete scale invariance [21,22], which
has recently been documented in the models of rupture, earthquake processes,
financial crashes, in the fractal geometry of growth processes, and in
several random systems [21-24]. Our approach can be straightforwardly
applied to series with complex powers. Then, the variable $s$ in the
algebraic transform (2) has also to be considered as complex. Therefore, the
corresponding control functions $s_k(x)$ become complex. However, before
passing to the general case of complex powers and, respectively, of complex
control functions, we need first to develop the approach for arbitrary real
powers and real control functions.

\section{Control Functions}

The role of control functions, by their definition [26], is to provide
convergence for initially divergent sequences. There are several ways of
incorporating control functions into an iterative algorithm and of defining
them. One way of introducing these functions is by including them into an
initial approximation. The most often used definitions of control functions
are through the minimal--difference [26-28] or minimal--sensitivity [29-36]
conditions. In some very simple cases, like zero-dimensional and
one-dimensional oscillators, when high--order terms of perturbation theory
are available explicitly, control functions can be found from the direct
observation of convergence of this theory [37-40]. All these variants are
particular types of quasifixed--point conditions [7-11].

Another way of introducing control functions is through an algebraic
transformation [12-14]. In addition, the minimal time appearing in the
evolution integral (7) under each renormalization step plays also the role
of a control function. After $k$ steps of the renormalization procedure, the
self--similar approximation (20) contains $k$ time--like control functions
$\tau _n,\; n=1,2,\ldots k$.

The simplest way of defining these control functions would be to
remember that the effective time in integral (7) corresponds to the minimal
number of steps needed for reaching a fixed point. When no other
restrictions are imposed, the minimal number of steps is, clearly, equal
to one. Putting $\tau _n=1$ for all $n=1,2,\ldots k$ in Eq.(20) gives 
\begin{equation}
\label{21}f_k^{*}(x)=F_k(x,1,1,\ldots ,1) . 
\end{equation}
A more elaborate definition of the time--like control functions can be
formulated
by involving one of the variants of fixed-point conditions. To this end, let
us put $\tau _n=1$ for all $n=1,2,\ldots k-1,$ except the last step for which 
$\tau _n=\tau$ is yet undefined. Consider the sequence $\{f_k\}$ consisting
of the terms 
\begin{equation}
\label{22}f_k(x,\tau )\equiv F_k(x,1,1,...,\tau ). 
\end{equation}
Following the standard procedure [10-14], it is possible to construct an
approximation cascade with a trajectory bijective to the sequence $\{f_k\}$. 
For the cascade velocity we may write the finite difference
\begin{equation}
\label{23}V_k(x,\tau )=f_k(x,\tau )-f_{k-1}(x,\tau ). 
\end{equation}
When approaching a fixed point, the cascade velocity tends to zero.
Therefore, the condition to be as close to the fixed point as possible is
the minimum of velocity (23). This condition 
\begin{equation}
\label{24}\min _\tau \left| V_k(x,\tau )\right| =\left| V_k(x,\tau_k(x))\right| 
\end{equation}
provides us the definition of the time-control function $\tau _k(x)$. Eq.(24)
is the general form of the minimal--velocity condition [12,41]. In
particular, taking into account definition (23), we may have the equation 
\begin{equation}
\label{25}f_k(x,\tau )-f_{k-1}(x,\tau )=0, 
\end{equation}
whose solution $\tau =\tau _k(x)$ is the sought control function. For the
exponential approximants (18), Eq. (25) yields the equation 
\begin{equation}
\label{26}\tau =\exp \left( \frac{a_k}{a_{k-1}}\tau \ x^{\beta _k}\right) . 
\end{equation}
The solution to Eq. (26), that is, $\tau =\tau _k(x),$ being substituted
into (22), leads us to the self--similar approximation 
\begin{equation}
\label{27}f_k^{*}(x)=f_k(x,\tau _k(x)). 
\end{equation}

The described scheme of defining control functions is applicable when no
additional restrictions are imposed on the behavior of the sought function 
$f(x)$. It may happen, that, in addition to expansion (1), an asymptotic
behavior of $f(x)$, as $x\rightarrow x_0,$ is known. Then the asymptotic
condition 
\begin{equation}
\label{28}f(x)\simeq f_0(x),\qquad x\rightarrow x_0 
\end{equation}
plays the role of an imposed restriction. And the constructed self--similar
approximations are assumed to satisfy the asymptotic condition (28). In such
circumstances, some of control functions are to be chosen so that condition 
(28) be valid. This can be done in the following way.

Let us renormalize a series $p_{k-1}(x)$ to a self--similar approximation 
$f^*_{k-1}(x)$ with time--like control functions defined according to a
scheme described above. Limiting ourselves by such a $(k-1)$--step
renormalization, we obtain from Eq.(10) 
\begin{equation}
\label{29}F_k^{*}(x,s,\tau )=\left[ \left( f^*_{k-1}(x)\right) ^{-\nu 
}-\frac{\nu \ a_k\tau }{a_0^{1+\nu }}x^{s\nu }\right] ^{-1/\nu }, 
\end{equation}
with the same notation (9). Now we require that the obtained expression (29)
would satisfy the asymptotic condition 
\begin{equation}
\label{30}F_k^{*}(x,s,\tau )\simeq f_0(x),\qquad x\rightarrow x_{0} ,  
\end{equation}
in accordance with (28). The control functions $s=s_k(x)$ and $\tau =
\tau_k(x)$ are to be chosen so that condition (30) be valid. With the so 
defined control functions, we come to the self-similar approximation 
\begin{equation}
\label{31}f_k^*(x)=F_k^{*}(x,s_k(x),\tau _k(x)) 
\end{equation}
possessing the desired asymptotic property (28).

To clearly illustrate the latter variant of defining control functions,
consider the integral 
\begin{equation}
\label{32}J(g)=\frac 1{\sqrt{\pi }}\int_{-\infty }^\infty \exp \left(
-x^2-gx^4\right) dx ,
\end{equation}
which has the meaning of the partition function for the zero--dimensional
anharmonic model, where g is called a coupling parameter. The series for
integral (32) are frequently used as a model for strongly divergent
perturbation expansions in quantum field theory [25].

The weak--coupling expansion of (32) gives 
\begin{equation}
\label{33}J(g)\simeq a+bg+...,\qquad (g\rightarrow 0), 
\end{equation}
with $a=1,\ b=-\frac 34.$ In the strong-coupling limit one has 
\begin{equation}
\label{34}J(g)\simeq Ag^{-1/4}+Bg^{-3/4}+Cg^{-5/4}+Dg^{-7/4}...,\qquad
(g\rightarrow \infty ), 
\end{equation}
where
$$
A=\frac{1.813}{\sqrt{\pi }},\qquad B= -\frac{0.612}{\sqrt{\pi }},\qquad 
C=\frac{0.227}{ \sqrt{\pi }}. 
$$
The strong--coupling expansion (34) is of the form of series (1) with the
coefficients
$$
a_0=A,\qquad a_1=B,\qquad a_2=C,\qquad a_3=D 
$$
and with noninteger powers 
$$
\alpha _0=-\frac 14,\qquad \alpha _1=-\frac 34,\qquad \alpha _2=-\frac 54,
\qquad \alpha _3=-\frac 74. 
$$
We shall renormalize expansion (34) subject to the limiting condition

\begin{equation}
\label{35}\lim _{g\rightarrow 0}J(g)=a, 
\end{equation}
which follows from (33).

Starting from 
\begin{equation}
\label{36}J_0(g)=Ag^{-1/4}, 
\end{equation}
we get 
\begin{equation}
\label{37}J_2^{*}(g)=J_0(g)\exp \left( \frac BAg^{-1/2}\exp \left( \frac
CBg^{-1/2}\right) \right) . 
\end{equation}
At the next step, according to (29), we have 
\begin{equation}
\label{38}F_3^{*}(g,s,\tau )=\left[ \left( J_2^{*}(g)\right) ^{-\nu }+\gamma
\ g^{s\nu }\right] ^{-1/\nu }, 
\end{equation}
with
$$
\nu =\frac 6{1-4s} ,\qquad \gamma =-\frac{\nu \ D\tau }{A^{1+\nu }}. 
$$
In the weak--coupling limit, Eq. (37) yields
$$
J_2^{*}(g)\simeq Ag^{-1/4}\quad (g\rightarrow 0) 
$$
and Eq. (38) gives
$$
F_3^{*}(g,s,\tau )=\left( A^{-\nu }g^{\nu /4}+\gamma \ g^{s\nu }\right)
^{-1/\nu }. 
$$
The latter expression satisfies the limiting condition 
\begin{equation}
\label{39}\lim _{g\rightarrow 0}F_3^{*}(g,s,\tau )=a, 
\end{equation}
corresponding to (35), if and only if $s=0$ and $\gamma =a\ $or,
respectively,
$$
\nu =6,\qquad \tau =-\frac{A^7}{6a^6D}\ . 
$$
Substituting the found s and $\tau $ into (38), we get%
$$
J_3^{*}(g)=\left[ \left( J_2^{*}(g)\right) ^{-6}+a^{-6}\right] ^{-1/6}. 
$$
Using here form (37) and remembering that $a=1$, we finally obtain the
self--similar approximation 
\begin{equation}
\label{40}J_3^{*}(g)=\left[ 1+\frac{g^{3/2}}{A^6}\exp \left( \frac{-6B}{A 
\sqrt{g}}\exp \left( \frac C{B\sqrt{g}}\right) \right) \right] ^{-1/6}
\end{equation}
corresponding to (31).

The exponential approximant (40) represents integral (32) with a very good
accuracy. The maximal percentage error is $2.7\%$ occurring at $g=0.1$. And
the error for the physically most interesting region of intermediate 
$g\approx 1$ is only $0.7\%.$

Consider another simple example evidently illustrating the generality of 
our method that works when other methods do not. Assume that we need to 
find an approximation for a bounded function $f(x)$, with 
$-\infty< x<\infty$, having the following asymptotic properties:
$$ f(x) \simeq 1 + x +O(x^2) \qquad (x\rightarrow 0) , $$
$$ f(x) \simeq e^x \qquad (x\rightarrow -\infty) , $$
$$ f(x)\simeq \frac{1}{x^\alpha} , \qquad \alpha > 0 \qquad 
(x\rightarrow +\infty) , $$
where $\alpha$ is an irrational power. To our understanding, Pad\'e technique
is principally inappropriate in this case. While in our method, 
accomplishing the same procedure as is explained above, we easily obtain
$$ f^*(x) =\left [ \exp\left (-\frac{2}{\alpha}x\right ) + 
x^2\right ]^{-\alpha/2} . $$

\section{Stability Conditions}

In order to check the stability of calculational procedure, one has to
analyze mapping multipliers [9-11]. Several kinds of multipliers occur in
the process of construction of self-similar approximations, each kind being
related to the corresponding approximation cascade.

The first approximation cascade, appearing in our investigation is the
cascade $\{y_k\}\ $composed of transforms (5). The local multipliers for
this cascade are defined as 
\begin{equation}
\label{41}\mu _k(\varphi ,s)\equiv \frac \partial {\partial \varphi
}y_k(\varphi ,s)=\sum_{n=0}^k\frac{\left( s+\alpha _n\right) a_n}{\left(
s+\alpha _0\right) a_0}\left( \frac \varphi {a_0}\right) ^{\left( \alpha
_n-\alpha _0\right) /\left( s+\alpha _0\right) }. 
\end{equation}
The image of multiplier (41) in the $x$--space can be obtained with the
use of relation (3), which yields 
\begin{equation}
\label{42}m_k(x,s)\equiv \mu _k(P_0(x,s),s)=
\frac{\partial P_k(x,s)/\partial x}{\partial P_0(x,s)/\partial x} =
\sum_{n=0}^k\frac{\left( s+\alpha _n\right) a_n}{\left( s+\alpha _0\right)
a_0}x^{\left( \alpha _n-\alpha _0\right) }. 
\end{equation}
The trajectory \{$y_k(\varphi ,s)$\} of an approximation cascade \{$y_k$\}
is locally stable [11] at the $k$-step if $\left| \mu _k(\varphi ,s)\right|
<1.$ Note that the case of $\left| \mu _k\right| =1$ is called neutrally
stable and that of $\left| \mu _k\right| =0$ can be termed superstable.
If $\left| \mu _k\right| <1$ for some $s$ and all $\varphi $ from a given
domain, then $\left| m_k(x,s)\right| <1$ for the same $s\ $and all $x$ from
a domain defined by relation (4). When the trajectory of a cascade 
\{$y_k $\} is locally stable for all $k=0,1,2,\ldots$, that is, when 
$\left| m_k(x,s)\right| <1$ for all $k$, then the sequence \{$P_k(x,s)$\} 
converges uniformly with respect to $x$. The local stability for all 
$k=0,1,2,\ldots$ can be called the global stability. The global stability of 
a trajectory is a {\it sufficient} condition for the convergence of the 
corresponding sequence [11]. But it is not a necessary condition. Thus, 
the sequence \{$P_k(x,s)$\} may be convergent, but the cascade trajectory 
not everywhere locally stable, so that the stability condition $\left| 
m_k(x,s)\right| <1$ becomes valid for all $k$ starting from some $k_0,$ 
but for $k<k_0$ this condition may be broken for some $k$.

The concept of stability suggests a recipe of defining the control function 
$s=s_k(x)$. The latter can be defined so that to minimize the absolute
value of multiplier (42), which can be named the principle of maximal
stability [12-14]. Substituting the found $s_k(x)$ into $P_k(x,s)$, we obtain 
$P_k(x,s_k(x)).$ The renormalized sequence \{$P_k(x,s_k(x))$\} can become
convergent even if the initial sequence \{$P_k(x,s)$\} was not. Defining
local multipliers for the new sequence, it is necessary to take into account
that $P_k(x,s_k(x))$ depends on $x$ explicitly as well as through 
$s_k(x)$. The corresponding multipliers are 
\begin{equation}
\label{43}m_k(x)=\frac{\partial P_k/\partial x+\left( \partial P_k/\partial
s_k\right) \left( ds_k/dx\right) }{\partial P_0/\partial x+\left( \partial
P_0/\partial s_k\right) \left( ds_k/dx\right) }\; ,
\end{equation}
where $P_k=P_k(x,\ s_k),\ s_k=s_k(x),$ and the partial derivatives mean that
another variable is kept fixed. In general, $m_k(x)$ differs from (42). 
But if $s_k(x)$ is a slowly varying function of $x$, such that the
derivative $ds_k/dx$ in $m_k(x)$ can be neglected, then $m_k(x)$
approximately coincides with $m_k(x,s_k(x)).$ The stability condition
$\left| m_k(x)\right| <1$ for all $k=0,1,2,\ldots$ implies the convergence
of the sequence $\{ P_k(x,s_k(x))\}$. Let us stress again that stability 
is a sufficient condition for convergence, but not necessary. The sequence
$\{ P_k(x,\ s_k(x))\}$ may converge even if the stability condition is not
valid for a finite value of $k$. Moreover, the convergence of the sequence
$\{ P_k(x,\ s_k(x))\}$ is not compulsory for us. This sequence is not yet
the final product of the procedure but is to undergone the dynamical
renormalization involving the evolution integral (7).

After the multiple dynamical renormalization, in the exponential variant we
consider here, we come to the sequence \{$F_k(x,\tau )$\} of terms given by
expression (18), with the shorthand notation $\tau \equiv \{\tau _1,\tau
_{2,}...,\tau _k\}$. The stability of the corresponding trajectory is
characterized by the multipliers 
\begin{equation}
\label{44}M_k(x,\tau )\equiv \frac{\partial F_k(x,\tau )/\partial x}{
\partial F_1(x,\tau )/\partial x}\; . 
\end{equation}
The stability condition $\left| M_k(x,\tau )\right| <1$ for all $k=1,2,\ldots$
guarantees the convergence of the sequence \{$F_k(x,\tau )$\}. This tells
us that the time-control functions $\tau _k(x)$ could be chosen so that to
minimize $\left| M_k(x,\tau )\right| .$ In particular, the minimal value of
(44) can be zero. Then the equality $M_k(x,\tau )=0,$ under the assumption
that $\partial F_1(x,\tau )/\partial x\neq 0,$ yields the equation 
\begin{equation}
\label{45}\frac \partial {\partial x}F_k(x,\tau )=0 
\end{equation}
restricting the choice of time-control functions.

Defining all time-control functions by one of the ways discussed above, we
obtain, as a final result, the exponential approximant (20). The sequence \{$%
f_k^{*}(x)$\}, with $k=1,2,...$, of these exponential approximants is what
we ultimately need to analyze with respect to its convergence. For that
purpose we can formulate sufficient conditions for convergence studying the
stability of the corresponding cascade. To this end, following the standard
procedure [7-11] , we define a function $x$($\varphi )$ by the equation 
\begin{equation}
\label{46}f_1^{*}(x)=\varphi ,\quad x=x(\varphi ). 
\end{equation}
Then we introduce the transformation 
\begin{equation}
\label{47}y_k^{*}(\varphi )=f_k^{*}(x(\varphi )). 
\end{equation}
The family \{$y_k^{*}$\} of the transformations introduced in (47) forms a
cascade. The local multipliers are given by 
\begin{equation}
\label{48}\mu _k^{*}(\varphi )\equiv \frac \partial {\partial \varphi
}y_k^{*}(\varphi ). 
\end{equation}
The image of (48) in the $x$-representation is 
\begin{equation}
\label{49}m_k^{*}(x)\equiv \mu _k^{*}(f_1^{*}(x))=\frac{\partial
f_k^{*}(x)/\partial x}{\partial f_1^{*}(x)/\partial x}. 
\end{equation}
The cascade trajectory is locally stable at a $k$-step when 
\begin{equation}
\label{50}\left| m_k^{*}(x)\right| <1. 
\end{equation}
If this condition is valid for all $k$, then the sequence \{$f_k^{*}(x)$\} 
converges.

The validity of the stability condition (50) is what finally justifies the
whole renormalization procedure guaranteeing the convergence of the
renormalized sequence \{$f_k^{*}(x)$\}. This holds true irrespectively of
whether other multipliers, as (42), (43), or (44), satisfy the same
condition (50) as the multiplier (49). The auxiliary multipliers (42)-(44)
have appeared at intermediate stages of our multistep renormalization
procedure. The minimization of these multipliers provides a recipe for
defining control functions. This minimization is equivalent to the
stabilization of local parts of a cascade trajectory. Remind that we start
with series (1) which, in general, is strongly divergent. Accomplishing
several stages of renormalization we step by step improve convergence
properties, by making the related trajectory more and more
stabilized. However, it is not necessary to demand that the trajectory would
become completely stable at some intermediate step of the multistep
procedure. The main is that the finally resulting sequence \{$f_k^{*}(x)$\}
be convergent, for which it is sufficient to require the validity of
condition (50).

\section{ Ground State Properties}

Now we pass to the consideration of physical examples illustrating 
explicitly how the method works.

\subsection{One-Dimensional Bose System}

The ground-state energy of the one-dimensional Bose system with the 
$\delta$--functional repulsive interaction potential is known in a 
numerical form from the Lieb--Liniger exact solution [42]. We derive below 
a compact analytical expression for the ground--state energy $e(g)$ as a 
function of the interaction strength $g$, valid for arbitrary $g$. In the 
weak--coupling and strong--coupling limits the following expansions are 
known (see e.g. Ref.[7] and references therein) : 
\begin{equation}
\label{51}e(g)\simeq ag+bg^{3/2}+cg^2+dg^{5/2}+\ldots \ ,\qquad 
\left( g\rightarrow 0\right) , 
\end{equation}
$$
a=1,\qquad b=-\frac 4{3\pi },\qquad c=0.0654,\qquad d=-0.0018, 
$$
while in the strong-coupling limit an exact result is available [43]: 
\begin{equation}
\label{52}e(g)\simeq A=\frac{\pi ^2}3\ ,\qquad \left( g\rightarrow \infty
\right) . 
\end{equation}
The expression satisfying both known limits, can be derived similarly to the
example studied in Section III, except that in this case we start from the
weak-coupling limit. The resulting approximant is 
\begin{equation}
\label{53}e_4^{*}(g)=ag\left[ \left( \exp \left( \frac bag^{1/2}\exp \left(
\frac cbg^{1/2}\right) \right) \right) ^{-3/2}+\left( \frac Aa\right)
^{-3/2}g^{3/2}\right] ^{-2/3}. 
\end{equation}
This expression works reasonably well in the region of $g\in [0,10]$, 
where the exact numerical solution of the Bethe ansatz equations is
available. The maximal error here is about $5\%$.

\subsection{Asymmetric Anderson Hamiltonian}

The single--orbital Anderson Hamiltonian describes the system consisting of
localized $d$--electrons interacting with conducting $s-$electrons via
a quantum--mechanical exchange mechanism, whose strength is measured by $V$,
which is the transfer integral between the $s$-- and $d$--states. The 
energy $U$ describing the Coulomb interactions between two $d$--electrons
is another relevant physical parameter. The localized $d-$electron number 
$n_d$ is expressed through the self--energy $\Sigma$ of $d$--electrons at 
the Fermi level: 
\begin{equation}
\label{54}n_d=\frac 12-\frac 1\pi \tanh {}^{-1}\left( \frac \Sigma \Delta
\right) ,\qquad \Delta =\pi \rho V^2, 
\end{equation}
where $\rho $ stands for the density of states at the Fermi level for
conducting electrons. In the case when $d$--level is fixed to the Fermi--level,
the following expansion for $\Sigma ,\ $in powers of the small parameter 
$u=U/\pi \Delta ,\left| u\right| \ll 1\ ,\ $was obtained in Ref. [44]:
\begin{equation}
\label{55}\frac \Sigma \Delta \simeq \frac \pi 2u\left(
1+a_1u+a_2u^2+a_3u^3+a_4u^4+a_5u^5\right)  \qquad
(u\rightarrow 0) ,
\end{equation}
$$
a_1=-1,\qquad a_2=0.5326,\qquad a_3=0.6269,\qquad a_4=-1.8071,
\qquad  a_5=1.027. 
$$

A direct application of expansion (55), gives physically meaningful results
in the region of $-0.5<u<0.5$, as is shown in Fig.2 of Ref.[44].
However, the exactly known limits 
$$
n_d\rightarrow 0\quad (u\rightarrow \infty );\qquad n_d\rightarrow 1\quad
(u\rightarrow -\infty ), 
$$
are violated when expansion (55) is used for the calculation of the impurity
level occupancy (54). To our knowledge, the Pad\'e approximants 
were not applied for resumming expansion (55), and it is not clear
whether they can be applied at large values of the parameter $u$ (see [2,45]).
We use below the technique of exponential approximants and observe that they
spontaneously recover the known limits. The following sequence of
self--similar renormalized expressions can be written: 
\begin{equation}
\label{56}\left( \frac \Sigma \Delta \right) _1^{*}=\frac \pi 2u\left( \exp
(a_1u)\right) ,
\end{equation}
\begin{equation}
\label{57}\left( \frac \Sigma \Delta \right) _2^{*}=\frac \pi 2u\left[ \exp
\left( a_1u\exp \left( \frac{a_2}{a_1}u\right) \right) \right] ,
\end{equation}
\begin{equation}
\label{58}\left( \frac \Sigma \Delta \right) _3^{*}=\frac \pi 2u\left[ \exp
\left( a_1u\exp \left( \frac{a_2}{a_1}u\exp \left( \frac{a_3}{a_2}u\right)
\right) \right) \right] ,
\end{equation}
\begin{equation}
\label{59}\left( \frac \Sigma \Delta \right) _4^{*}=\frac \pi 2u\left[ \exp
\left( a_1u\exp \left( \frac{a_2}{a_1}u\exp \left( \frac{a_3}{a_2}u\exp
\left( \frac{a_4}{a_3}u\right) \right) \right) \right) \right] ,
\end{equation}
\begin{equation}
\label{60}\left( \frac \Sigma \Delta \right) _5^{*}=\frac \pi 2u\left[ \exp
\left( a_1u\exp \left( \frac{a_2}{a_1}u\exp \left( \frac{a_3}{a_2}u\exp
\left( \frac{a_4}{a_3}u\exp \left( \frac{a_5}{a_4}u\right) \right) \right)
\right) \right) \right] .
\end{equation}
We observe, that already the third--order approximant $\left( \frac \Sigma
\Delta \right )_3^{*}$ leads to the result   
\begin{equation}
\label{61}\left( n_d\right) _3^{*}=\frac 12-\frac 1\pi \tanh {}^{-1}\left[
\left( \frac \Sigma \Delta \right) _3^{*}\right] ,
\end{equation}
which possesses the correct limits. The same is true when the higher
order approximants
$$
\left( n_d\right) _4^{*}=\frac 12-\frac 1\pi \tanh {}^{-1}\left[ \left(
\frac \Sigma \Delta \right) _4^{*}\right] ,\qquad \ \left( n_d\right)
_5^{*}=\frac 12-\frac 1\pi \tanh {}^{-1}\left[ \left( \frac \Sigma \Delta
\right) _5^{*}\right] , 
$$
are considered. An aposteriori analysis of multipliers suggests that the
third-order approximant corresponds to the most stable trajectory. This
could be expected beforehand, since the coefficients $a_4\ $and $a_5$ are
larger than one. 

The density of states of $d$--electrons as well as the
resistivity $R$ at zero temperature, could be also reconstructed for 
arbitrary $u$, e.g. 
$$
\left( \frac{R(u)}{R(0)}\right) ^{*}=\left[ 1+\left( \left( \frac \Sigma
\Delta \right) _3^{*}\right) ^2\right] ^{-1}. 
$$
The shape of the curve $\left( n_d(u)\right) _3^{*}$ is very much like the
smeared Fermi--distribution, while $\left( \frac{R(u)}{R(0)}\right) ^{*}$ has
an asymmetric bell--shape. All artifacts, which appear at the curves, when
the perturbative expansions are naively extended beyond the region of small 
$u$, are smeared out, and the renormalized curves appear to be rather smooth.

\subsection{t--expansion}

The so--called  $t-$expansion is a tool for a systematic improvement of 
variational calculations for Hamiltonian systems [46]. Using a 
$t-$dependent variational wave function and, after performing calculations, 
finally, taking the $t\rightarrow\infty$ limit, one can hope to increase the 
quality of the variational estimate for the ground--state energy $E$. This 
idea is most frequently used in conjunction with heavy numerical calculations 
and Pad\'e--approximants technique. In the case of the $1d$ Heisenberg 
antiferromagnet, the expansion in powers of the parameter $t$ was obtained 
explicitly [46], 
\begin{equation}
\label{62}E \simeq -\frac{1}{4} - t+2t^2+\frac{4}{3} t^3-16t^4  \qquad 
(t\rightarrow 0) , \end{equation}
and the task of getting an estimate for the ground state energy from the
asymptotic expression valid at $t\rightarrow 0$, can be approached by the
methods of Sections II--IV. Since the number of terms available is finite, it
is not necessary to take the limit $t\rightarrow \infty $ explicitly, but
instead we can try to minimize the error caused by this inevitable
truncation by demanding the minimal sensitivity of the renormalized
expression $E^{*}$ with respect to the "time" $t$ determining the 
"duration" of motion to the ground state energy,
\begin{equation}
\label{63}\frac{\partial E^{*}}{\partial t}=0.
\end{equation}
Applying the self--similar bootstrap in its super--exponential form, one can
see that the solution to this equation does exist for any number of terms
from the initial expansion (62). From the {\it aposteriori}
analysis of the multipliers, we conclude that the most reliable value,
corresponding to that obtained along the most stable trajectory, is 
\begin{equation}
\label{64}E^{*}=-\frac{1}{4}\exp\left ( 4t\exp\left ( -2t\exp\left (
\frac{2}{3}t\right )\right )\right ),
\end{equation}
with $t=0.33$ and $E^{*}=-0.446$. From the viewpoint of an apriori
analysis, it is admissible also to construct another sequence of
exponentials, not including the constant, mean--field result. In this case,
the answer for the ground state energy is $-0.434$. We conclude that moving
along the two different, but stable, trajectories, we can determine $E$
reliably, as $E=-0.44\pm 0.006$. The known Hulthen [47] exact result
$-0.4431$ is located within these boundaries.

\section{Effective Coupling}

\subsection{Beta-Function of SU(2) Lattice Gauge Model}

The Callan--Symanzik $\beta -$function of the (3+1)--dimensional SU(2) 
lattice gauge model in its weak--coupling asymptotically free regime may be
presented in the form of an expansion in powers of the parameter $g,$ where 
$g$ stands for the coupling [48]: 
\begin{equation}
\label{65}-\frac{\beta (g)}g\simeq ag^2+bg^4+\ldots ,\qquad 
\left( g\rightarrow 0\right) , 
\end{equation}
$$
a=\frac{11}{24\pi ^2}\; ,\qquad b=\frac{17}{64\pi ^4}\; . 
$$
In its strong-coupling limit, $\beta $ can be presented as follows [48]: 
\begin{equation}
\label{66}-\frac{\beta (g)}g\simeq A+Bx^2+Cx^4+Dx^6+Fx^8 , \qquad 
x=\frac{4}{g^4}\qquad \left( g\rightarrow \infty \right) , 
\end{equation}
$$
A=1,\qquad B=-\frac{76}{75}, \qquad C=\frac{88}{625},\qquad 
D=-\frac{131203}{140625}, \qquad  F=\frac{551378}{390625}\; . 
$$
We will add to the expansion (66) one more trial term $\simeq Gx^{10},$ and
determine $G$ from the boundary condition following from  the
weak--coupling limit (65). Two starting terms from (66) will be
renormalized to the form of an exponential approximant, mimicking an 
instanton contribution bridging these two limits [49], that is, a
nonperturbative physical mechanism coming into play in the crossover
region, from strong--to--weak--coupling limit [50]. This term should
disappear completely in the weak--coupling limit, guarantying rather sharp
crossover. The last four terms will be renormalized to the form satisfying
the boundary condition (65), in a way leading to a smooth matching of two
limiting kinds of behavior. We obtain the following analytical form for
the renormalized $\beta -$function: 
\begin{equation}
\label{67} - \frac{\beta^*(g)}{g} = A \exp \left( \frac
BAx^2\right) +Cx^4\left( 1-\frac{\tau _1}{s_1}\frac DCx^2\right)
^{-s_1}+Fx^8\left( 1-\frac G{s_2F}x^2\right) ^{-s_2}, 
\end{equation}
$$
s_1=\frac 94,\quad 
\tau _1=-\frac{s_1C}D\left( \frac{2a}C\right)^{-1/s_1}=0.3;
\quad s_2=2s_1, \quad 
G=-s_2F\left( \frac{4b}F\right)^{-1/s_2}=-15.051. 
$$
The shape of the resulting function is similar to those obtained in
Refs.[48-50]. Crossover occurs at $g\sim 1.2$. Let us remind that
the quality of the strong-coupling expansions is jeopardized by the
interfering roughening transition [48]. We believe that the method 
discussed above, allows us to bypass this difficulty in a natural way.

\subsection{Kondo Effect}

We consider below an application of the exponential approximants to such an
interesting problem as the Kondo effect [51], comparing our results with
those of the field--theoretical renormalization group [52-54].

The behavior of the system, consisting of a local--impurity spin and conduction
electrons, interacting by means of an antiferromagnetic exchange of strength 
$J$, changes from asymptotically free at high temperatures to that when the
impurity is screened by electronic lump at low temperatures, via the 
crossover 
region whose onset is characterized by the Kondo temperature estimated as 
\begin{equation}
\label{68}T_k=D\exp \left( -\frac 1{2J}\right) , 
\end{equation}
where $D$ stands for the Fermi--energy of electrons. We consider below only
the case of a single--channel Kondo model. Most of our knowledge about the
problem came from the exact Bethe ansatz solution [55,56], from the
field--theoretical renormalization group [52-54], and from the Wilson
numerical renormalization group [57].

Within the framework of the field-theoretical RG in its application to the
Kondo crossover, the central role is played by the so-called invariant
charge or effective electron-electron coupling$\ J_{inv}$ [52-54], measuring
the intensity of electron-electron interactions via the impurity spin: 
\begin{equation}
\label{69}J_{inv}=J\left[ 1+2J\ln \left( \frac D{\left| \omega \right|
}\right) -2J^2\ln \left( \frac D{\left| \omega \right| }\right) +
\ldots \right] ,
\end{equation}
here $\omega $ stands for the typical external parameter of the problem,
such as temperature, or magnetic field. If only the starting two terms from
(69) are taken into account, the field--theoretical approach, through
calculations based on the Gell--Mann--Low $\beta -$function, 
\begin{equation}
\label{70}\beta \simeq -2J^2, \qquad (J\ll 1),
\end{equation}
leads to the formally divergent, at $T=T_k$, expression for the invariant
charge [52-54],
\begin{equation}
\label{71}J_{inv}=\frac J{1-2J\ln \left( \frac D{\left| \omega \right|
}\right) }\ .
\end{equation}
We apply below the technique of algebraic self-similar renormalization {\it 
directly} to the series (69) for $J_{inv}$, continuing them from the region
of $J\ll 1$ to the region of $J\sim 1$. When only two starting terms from
(69) are taken into account, the optimal, from the viewpoint of stability 
conditions, solution is the following exponential approximant,
\begin{equation}
\label{72}J_{inv}^{*}=J\exp \left[ 2J\ln \left( \frac D{\left| \omega
\right| }\right) \right] =J\ \left( \frac D{\left| \omega \right| }\right)
^{2J}.
\end{equation}
If we perform the self--similar renormalization with the control function 
$s=0$, along the non--optimal trajectory, we will recover the expression 
(71). So, the fictitious pole is absent in our solution to the problem, 
although the typical energy scale determined from the condition
$$
2J\ln \left( \frac D{\left| \omega \right| }\right) \sim 1, 
$$
coincides with the Kondo temperature (68). Expression (72) is formally
divergent as $\omega \rightarrow 0,$ in agreement with the conclusion of the
numerical renormalization group [57,58].

More complicated situation arises in (quasi)--two--dimensional metal, when the
Van Hove logarithmic singularity in the electron density of states can
influence the Kondo effect. In this case, as was shown in Ref.[59], the
full electron--impurity scattering amplitude $\Gamma $ can be estimated 
as follows: 
\begin{equation}
\label{73}\Gamma =J\left[ 1+c_0J\ \left( \ln \left( \frac D{\left| \omega
\right| }\right) \right) ^2+\ldots \right] ,\qquad c_0>0,\qquad 
\left( J\ll 1\right) , 
\end{equation}
i.e. the usual Kondo logarithm should be replaced by the squared logarithm,
originating from the Van Hove singularity. The self--similarly
renormalized expression in this case again corresponds to the exponential
approximant: 
\begin{equation}
\label{74}\Gamma =J\ \exp \left[ c_0J\ \left( \ln \left( \frac D{\left|
\omega \right| }\right) \right) ^2\right] , 
\end{equation}
and the characteristic energy scale, Kondo temperature, can be found from
the condition
$$
c_0 J\ \left( \ln \left( \frac D{\left| \omega \right| }\right) \right) 
^2\sim 1, 
$$
leading to the estimate 
\begin{equation}
\label{75}T_k=D\exp \left( -\frac{const}{\sqrt{J}}\right) ,\qquad 
const\sim \sqrt{\frac 1{c_0}}. 
\end{equation}
Such a dependence of $T_k$ on $J$ was the main result of Ref.[59], 
obtained as an outcome of a cumbersome first order parquet summation .
To our knowledge, the field--theoretical RG approach was not applied to the
Kondo effect with the Van Hove singularity. On the other hand,
the Bethe ansatz fails for this problem [59].

The higher--order corrections were not considered in Ref.[59], because of the
technical difficulties arising in more sophisticated parquet approximations.
Our approach may be of interest in this context, allowing to find
the corrections to Kondo temperature due to higher--order
terms. Taking into account the higher--order perturbative terms, one has
\begin{equation}
\label{76}\Gamma =J\left[ 1+c_0J\ \left( \ln \left( \frac D{\left| \omega
\right| }\right) \right) ^2+c_1J^2\left( \ln \left( \frac D{\left| \omega
\right| }\right) \right) ^2+...\right] ,\quad c_0>0,\quad 
\left( J\ll 1\right) , 
\end{equation}
with $c_1<0$. The following exponential approximant is
optimal from the viewpoint of stability: 
\begin{equation}
\label{77}\Gamma =J\ \exp \left[ c_0J\ \left( \ln \left( \frac D{\left|
\omega \right| }\right) \right) ^2\exp \left( -\frac{\left| c_1\right| }{c_0}%
J\right) \right] .
\end{equation}
From the condition
$$
c_0 J\ \left( \ln \left( \frac D{\left| \omega \right| }\right) \right)^2
\exp \left( -\frac{\left| c_1\right| }{c_0}J\right) \sim 1, 
$$
we obtain the Kondo temperature: 
\begin{equation}
\label{78}T_k=D\exp \left[ -\frac 1{c_0^{1/2}J^{1/2}}\exp \left( \frac 12 
\frac{\left| c_1\right| }{c_0}\ J\right) \right] . 
\end{equation}
This estimate suggests a decrease of $T_k$ due to higher--order
corrections.

\section{Equation of State}

\subsection{ Classical Hard Spheres.}

The exponential approximants can be used for constructing equations of
state for simple liquids. For the model system of hard spheres with the
diameter $d$, widely used as a reference system, an empirical equation
of state, suggested by Carnagan and Starling (see [60]), is 
\begin{equation}
\label{79}\frac p{nkT}=\frac{1+\rho +\rho ^2-\rho ^3}{(1-\rho )^3}\; ,
\end{equation}
connecting pressure $p$, temperature $T$, the number density $n$, and the
reduced density $\rho =\pi nd^2/6$. It agrees very well with
the molecular dynamics and virial expansion [60,61]. The theoretical virial
formula, according to Percus--Yevick [60,61], is given as follows: 
\begin{equation}
\label{80}\frac p{nkT}=\frac{1+\rho +\rho ^2-3\rho ^3}{(1-\rho )^3}\ . 
\end{equation}
These two expressions almost coincide at low densities, e.g. at $\rho =0.1$
the percentage error of Eq.(80), as compared to (79), equals $-0.18\%$;
while for the intermediate and high densities the agreement becomes very
poor, e.g. at $\rho =0.5$ the percentage error is $-15.385\%$, and at $\rho
=0.7$ it equals $-37.141\%$ .

Consider the regular part of (80), defined as $r:$%
\begin{equation}
\label{81} r\simeq 1+\rho +\rho ^2-3\rho ^3 \qquad (\rho\rightarrow 0) ,
\end{equation}
as an asymptotic, low--density expansion for the regular part 
$r(\rho )$, and try to continue the expression (81) from the region of 
$\rho \ll 1$, to the region of $\rho \leq 1$.$\ $In order to extend the
validity of (81), let us add to it one more trial term $\sim \rho ^4$. It
is reasonable to use for renormalization the last four terms from thus
extended expansion for $r$, since the constant term describes the ideal gas
behavior and we are interested in the region of high densities. Following
the standard prescriptions of Sec.II--III, we write down the two exponential
approximants, justified from the viewpoint of stability for the sequence of
an aposteriori multipliers: 
\begin{equation}
\label{82}r_3^{*}(\rho ,\tau )=1+\rho \exp \left[ \rho \exp \left( -3\rho
\tau \right) \right] , 
\end{equation}
\begin{equation}
\label{83}r_4^{*}(\rho ,\tau )= 1 + \rho \exp \left[ \rho \exp \left(
-3\rho \exp \left( -\frac \tau 3\rho \right) \right) \right] . 
\end{equation}
We retained in expressions (82) and (83) the effective time $\tau $,
introduced at the last step of the bootstrap procedure. It will work now as
a control function $\tau =\tau (\rho )$ determined from the minimal--velocity 
condition $\min _\tau \left| r_4^{*}(\rho ,\tau )- 
r_3^{*}(\rho ,\tau )\right|$. For the sake of simplicity we choose a 
single control parameter $\tau$, instead of the control function, from 
the minimal difference condition imposed at a single point, chosen 
from the region of intermediate densities, say for $\rho =0.6$. 
Then, $\tau =0.845$. Recalculating 
\begin{equation}
\label{84}\frac{p^{*}}{nkT}=\frac{r_4^{*}(\rho ,\tau )}{(1-\rho )^3}, 
\end{equation}
and comparing it with empirical formula (79), we obtain that at $\rho
=0.1$ the percentage error equals $-0.112\%$ ; at $\rho =0.5$ the
percentage error is $-3.215\%$, and at $\rho =0.7$ it equals $-2.92\%.$
Formula (84) is more accurate at high densities than our
previous result [14], corresponding to 
\begin{equation}
\label{85}\frac{p^{*}}{nkT}=\frac{r_3^{*}(\rho ,1)}{(1-\rho )^3}, 
\end{equation}
which works with the percentage error of $-4.567\%$ at $\rho =0.7$. What is
even more important here, is the possibility of a self--consistent
improvement of the quality of the equation of state, based on stability
conditions. We should recall here, that phenomenological exponential--type
expressions are well known in the theory of equations of state, beginning,
probably, from the Hudleston equation (see [61] and references therein) and
ending with its modern modifications [62-64]. For example, the Shinomoto's
equation for the system of hard spheres reads: 
\begin{equation}
\label{86}\frac p{nkT}=\exp \left[ 4\rho \left( 1+\frac 12\rho \right)
\right] , 
\end{equation}
and gives at $\rho =0.1$ the percentage error $0.046\%$; at $\rho =0.5$ 
the error is $-6.289\%$, and at $\rho =0.7$ it becomes $-35.948\%$.

A single-exponential approximation can also be obtained from Eq.(81) by 
our method: 
\begin{equation}
\label{87}\frac{p^*}{nkT}=\frac{\rho \exp \left( \rho -3\rho
^2\right) +1}{(1-\rho )^3}, 
\end{equation}
giving at $\rho =0.1$ the error $-0.158\%$, at $\rho =0.5$ the error is 
$-14.498\%$, and at $\rho =0.7$ it becomes $-28.31\%$.

We conclude that the multi-exponential formula (84) agrees well with the
empirical formula (79), being superior to all other formulae in the region
of high densities.

\subsection{Quantum Hard Spheres}

At low density $\rho$, the energy $E\ $ for a boson system of $N$--hard
spheres with the diameter $c$ and mass $m$ is known [65]: 
\begin{equation}
\label{88}\frac EN\simeq \frac{2\pi }m\ \rho \ c\left( 1+C_1\left( \rho
c^3\right) ^{1/2}+...\right) ,\qquad C_1=\frac{128}{15\sqrt{\pi }}\qquad
\left( \rho \rightarrow 0\right) .
\end{equation}
And as $\rho \rightarrow \rho _0,$ where $\rho _0=\sqrt{2}/c^3$ is the
maximal density for a system of hard spheres, the following
expression is available [65]: 
\begin{equation}
\label{89}\frac EN\simeq A\frac 1{2m}\left( \rho ^{-1/3}-\rho
_0^{-1/3}\right) ^{-2},\qquad A=\frac{\pi ^2}{2^{1/3}}\qquad \left( \rho
\rightarrow \rho _0\right) , 
\end{equation}
corresponding to a second--order pole in the ground--state energy. Let us,
in analogy with the previous example, extract the singularity, as $\rho 
\rightarrow \rho _0$, rewriting (88) as follows: 
\begin{equation}
\label{90}\frac EN\simeq \frac{2\pi c}m\frac{\rho ^{1/3}\left[ 1+C_1\left(
\rho c^3\right) ^{1/2}+...\right] }{\left( \rho ^{-1/3}-\rho
_0^{-1/3}\right) ^2}\left[ 1-\left( \frac \rho {\rho _0}\right)
^{1/3}\right] ^2\ ,\qquad \left( \rho \rightarrow 0\right) , 
\end{equation}
and, keeping only a few starting terms 
\begin{equation}
\label{91}\frac EN\simeq \frac{2\pi c}m\frac{\rho ^{1/3}}{\left( \rho
^{-1/3}-\rho _0^{-1/3}\right) ^2}\left[ 1-2\left( \frac \rho {\rho
_0}\right) ^{1/3}+C_1\left( \rho c^3\right) ^{1/2}+...\right] .
\end{equation}
After the standard self--similar renormalization, involving two terms
from (91), we obtain 
\begin{equation}
\label{92} \frac{E^*}{N} =\frac{2\pi c}m\frac{\rho ^{1/3}}{%
\left( \rho ^{-1/3}-\rho _0^{-1/3}\right) ^2}\exp \left[ -2\tau \left( \frac
\rho {\rho _0}\right) ^{1/3}\right] , 
\end{equation}
where the control parameter $\tau$ should be determined from the known
asymptotic formula (89), as $\rho \rightarrow \rho _0.\ $Finally, 
\begin{equation}
\label{93} \frac{E^*}{N} =\frac{2\pi c}m\frac{\rho ^{1/3}}{
\left( \rho ^{-1/3}-\rho _0^{-1/3}\right) ^2}\exp \left[ \ln \left( \frac
A{4\pi c}\rho _0^{-1/3}\right) \left( \frac \rho {\rho _0}\right)
^{1/3}\right] ,\quad \tau =-\frac 12\ln \left( \frac A{4\pi c}\rho
_0^{-1/3}\right) , 
\end{equation}
or, equivalently, 
\begin{equation}
\label{94} \frac{E^*}{N}=\frac{2\pi c}m\frac{\rho ^{1/3}}{
\left( \rho ^{-1/3}-\rho _0^{-1/3}\right) ^2}\left( \frac A{4\pi c}\rho
_0^{-1/3}\right) ^{\left( \rho /\rho _0\right) ^{1/3}}. 
\end{equation}
Equations (93) and (94) should be compared with the empirical London equation
of state [66]. Both equations give the results very close to each other. So,
our derivation can serve as a justification for empirical formulae used for
the system of Bose hard spheres.

\subsection{Polymer Coil}

The expansion factor $\alpha $ of a polymer coil is
represented as a function $\alpha ^2=\alpha ^2(z)$ of the
excluded volume variable $z=B\sqrt{N}\left( 3/2\pi l^2\right) ^{3/2},$ where 
$N$ is the number of bonds of the length $l$ each and $B$ is an effective
binary cluster integral [67-70]. The case of a polymer coil corresponds to 
$z>0.$ In the region of $z\ll 1$, the perturbation theory in powers of $z$
can be developed, giving the expansion [68] 
\begin{equation}
\label{95}\alpha ^2=\alpha^2(z) \simeq
1+k_1z+k_2z^2+k_3z^3+k_4z^4+k_5z^5+k_6z^6 \qquad  (z\rightarrow 0) ,
\end{equation}
$$
k_1 =\frac 43,\qquad \ k_2=-2.075385396,\qquad k_3=6.296879676, 
$$
$$
k_4=-25.05725072,\qquad k_5=116.134785,\qquad k_6=-594.71663. 
$$
On the other hand, in the limit of $z\gg 1,$ $\alpha $ is related to $z$ by
a simple power law 
\begin{equation}
\label{96}\alpha ^2\simeq Kz^b,\qquad b=2(2\nu -1) \qquad 
(z\rightarrow\infty) ,
\end{equation}
where $\nu$ is the critical index, and $K$ stands for the critical
amplitude. One of the popular problems in the physics of polymer coils
consists in the continuation of the expansion (95) to the region of
arbitrary $z$. We derive below, using the self--similar renormalization, a
simple equation of state for the polymer coil, valid for arbitrary $z$, and
satisfying by design both known limits, (95) and (96). The coefficients in
expansion (95), starting from $k_3,$ grow rapidly, so do the local
multipliers. Because of this, we use only three starting terms from 
(95), stabilizing the renormalized expression by imposing the asymptotic 
condition (96). Finally, after the standard transformations analogous to 
those of Section III, we obtain 
\begin{equation}
\label{97}\left( \alpha ^2(z)\right) ^{*}=\left[ \left( \exp \left( k_1z\exp
\left( \frac{k_2}{k_1}z\right) \right) \right) ^{3/b}+K^{3/b}z^3\right]
^{b/3}. 
\end{equation}
It is known, from different approaches [12,69,70], that $1/2\leq $$\nu \leq 
0.6$ and $1.53\leq $$K\leq 1.75$. We take the values $\nu=0.599$ and $K=1.62$, 
which we calculated in Ref.[12],  and compare the equation of state (97) 
with the empirical equations of state [70] of Barrett--Domb,
\begin{equation}
\label{98}\alpha ^2=\left( 1+\frac{20}3z+4\pi z^2\right) ^{1/5}, 
\end{equation}
and of Yamakawa--Tanaka, 
\begin{equation}
\label{99}\alpha ^2=0.572+0.428\left( 1+6.23z\right) ^{1/2}. 
\end{equation}
In the region $0\leq z\leq 12$,  our equation (97) yields a curve lying 
between those of eqs.(98) and (99).

\subsection{Ising Model}

The low--temperature expansion for the order parameter (magnetization) $M$
of the three--dimensional Ising model on a f.c.c. lattice reads [71]: 
\begin{equation}
\label{100}M(T)\simeq 1+bu^6+cu^{11} , \qquad u=\exp \left( -\frac 4T\right)
\qquad (T\rightarrow 0),
\end{equation}
$$
b=-2,\qquad c=-24. 
$$
Low--temperature expansions are non--universal in the sense that they 
depend on the type of a lattice, spin etc$\ldots$. At the critical point
$T_c$, order parameter demonstrates a universal behavior, independent on 
the type of a lattice, spin etc$\ldots$: 
\begin{equation}
\label{101}M\sim \left( T_c-T\right) ^\beta ,\qquad \left( T\rightarrow
T_c\right) ,
\end{equation}
where $\beta \approx 0.325$ is the critical index [72]. For the f.c.c.
lattice, $T_c\approx 9.8$ [72]. We continue the low--temperature expansion
(100) self--similarly to the whole region of $0\leq $$T\leq T_c,$ imposing
relation (101) as a boundary condition. Following the standard prescriptions
of Section III, we obtain 
\begin{equation}
\label{102}M^{*}(T)=\left[ \left( \exp \left( bu^6\right) \right) ^{-11/s}-
\frac{11c}s\tau u^{11}\right] ^{-s/11},
\end{equation}
$$
s=-11\beta ,\qquad \tau =-\frac{\beta}{cu^{11}(T_c)}\exp \left( 
\frac b\beta \ u^6(T_c)\right) =0.718. 
$$
In the region of intermediate temperatures, Eq. (102) agrees much better 
with the experimental data for the magnetization of $Fe$ and $Ni$ [73] than
the Bragg--Williams approximation. In distinction from the Burley
extrapolation [74], our formula is simple and has a transparent physical
background, taking into account both short--and--middle--range correlations, 
which contribute through the exponential function at low and intermediate
temperatures, as well as long--range correlations dominating at the critical
point.

\section{Magnetic Properties}

The ground state properties of the two--dimensional Heisenberg
antiferromagnet can be considered by means of an expansion around its 
Ising limit in powers of the anisotropy parameter $x$, equal to zero 
for the Ising model and equal to one for the Heisenberg--limit [75,76]. Such 
expansions can be generated separately for the most interesting cases of 
spin--$1/2$ and spin--$1$, allowing thus to treat them independently, not 
relying on the expansion in inverse powers of spin around a very distant 
case of classical spins.

The expression for susceptibility $\chi $ of the two-dimensional Heisenberg
antiferromagnet was obtained in the following form [75,76]:
\begin{equation}
\label{103}\frac{1}{2}\chi \simeq \frac 18+a_1x+a_2x^2+a_3x^3+a_4x^4 \qquad
(x\rightarrow 0) ,
\end{equation}
$$
a_1=-\frac 16,\quad a_2=0.177083,\quad a_3=-0.1898148,\quad
a_4=0.191761,\quad a_5=-0.196579, 
$$
$$
a_6=0.197934,\qquad a_7=-0.201447\qquad \left( S=\frac 12\right) ;
$$
$$
a_1=-0.142857,\quad a_2=0.144643,\quad a_3=-0.149916,\quad
a_4=0.150672,\quad a_5=-0.153095, 
$$
$$
a_6=0.153437,\qquad a_7=-0.154932\qquad \left( S=1\right) . 
$$
According to general prescriptions of Section IV, we should, first, analyze
the values of local multipliers $m_k(s)$, as $s\rightarrow \infty $. 

Consider the case of $S=1$. Since the values of the coefficients in the 
expansion (103) are slowly growing and oscillate with increasing number, 
the local multipliers will oscillate too. We conclude that in some cases 
we can continue the trajectory moving along the stable regions, while in 
other cases we have to move along unstable regions. In this situation we must
rely on the aposteriori analysis. In order to choose the starting term in 
(103), let us compare the values of $m_k(s)$, as $s\rightarrow\infty$. Then  
$m_1(\infty )=1+a_2/a_1=-0.143$ and $m_2(\infty)=1+a_3/a_2=-0.013$. 
Since the latter number is smaller, we should start the renormalization 
procedure from the linear term, keeping the constant term untouched. The 
following sequence of exponential approximants can be readily
written down, with the corresponding multipliers shown in brackets, 
calculated at $\tau=1$ and $x=1$: 
\begin{equation}
\label{104}\frac 12\ \chi _2^{*}(\tau )=\frac 18+a_1x\exp \left( \frac{a_2}
{a_1}x\tau \right) =0.146 \qquad (m_1^{*}=1),
\end{equation}
\begin{equation}
\label{105}\frac 12\ \chi _3^{*}(\tau )=\frac 18+a_1x\exp \left[ \frac{a_2}
{a_1}x\exp \left( \frac{a_3}{a_2}x\tau \right) \right] =0.05
\qquad (m_2^{*}=-0.025),
\end{equation}
\begin{equation}
\label{106}\frac 12\ \chi _4^{*}(\tau )=\frac 18+a_1x\exp \left[ \frac{a_2}
{a_1}x\exp \left( \frac{a_3}{a_2}x\exp \left( \frac{a_4}{a_3}x\tau \right)
\right) \right] =0.107 \quad (m_3^{*}=0.944),
\end{equation}
$$
\frac 12\ \chi _5^{*}(\tau )=\frac 18+a_1x\exp \left[ \frac{a_2}{%
a_1}x\exp \left( \frac{a_3}{a_2}x\exp \left( \frac{a_4}{a_3}x\exp \left( 
\frac{a_5}{a_4}x\tau \right) \right) \right) \right] =0.075
$$
\begin{equation}
\label{107} (m_4^{*}=0.225), 
\end{equation}
$$
\frac 12\ \chi _6^{*}(\tau )=\frac 18+a_1x\exp \left[ \frac{a_2}{
a_1}x\exp \left( \frac{a_3}{a_2}x\exp \left( \frac{a_4}{a_3}x\exp \left( 
\frac{a_5}{a_4}x\exp \left( \frac{a_6}{a_5}x\tau \right) \right) \right)
\right) \right] =0.094~
$$
\begin{equation}
\label{108}(m_5^{*}=0.754), 
\end{equation}
$$
\frac 12\ \chi _7^{*}(\tau )=\frac 18+ $$
$$ + a_1x\exp \left[ \frac{a_2}{
a_1}x\exp \left( \frac{a_3}{a_2}x\exp \left( \frac{a_4}{a_3}x\exp \left( 
\frac{a_5}{a_4}x\exp \left( \frac{a_6}{a_5}x\exp \left( \frac{a_7}{a_6}x\tau
\right) \right) \right) \right) \right) \right] =0.083~
$$
\begin{equation}
\label{109}\left( m_6^{*}=0.394\right) . 
\end{equation}
We observe two subsequences, with odd and even numbers, probably
embracing the correct result from below and above, respectively. We can
suspect that they both define a stable quasifixed point corresponding to a
focus. In order to locate it with maximal possible precision, we can impose
a minimal difference condition on the points, belonging one to the "odd"
and another to "even" subsequences, with the smallest absolute values of
multipliers. On the other hand, it is interesting to compare the values
that can be obtained from the minimal--difference condition imposed on
two starting terms of the renormalized sequence with those obtained from 
two last terms. E.g., from the condition 
$\min _\tau \left| \chi _3^{*}(\tau )-\chi _2^{*}(\tau )\right|$,
we obtain the value of the control parameter $\tau =0.56$. Correspondingly, 
$\chi _3^{*}(\tau ,S=1)=0.088.$ We obtain thus, $0.087\leq\chi^*(\tau,S=1)
\leq 0.088$, with the lower bound following from the analysis of two last
terms and of two approximants with minimal multipliers. Our estimate
should be compared to the result of Ref. [76], $0.095\pm 0.002$, obtained
by Pad\'e--summation.

The case of $S=\frac 12$ may be treated by analogy with the case of $S=1$.
Following literally the same steps as above, we obtain 
$0.062\leq \chi^*(\tau,S=\frac{1}{2})\leq 0.064,$ being close to the
result of Pad\'e--summation, $0.065\pm 0.003$ [75].

The ground state energy $E$ of the two--dimensional Heisenberg
antiferromagnet may be presented in the form of an expansion [75,76]: 
\begin{equation}
\label{110}2E \simeq -4+a_2x^2+a_4x^4+a_6x^6+a_8x^8 \qquad (x\rightarrow 
0) ,
\end{equation}
$$
a_2=-\frac 43,\quad a_4=0.0064,\quad a_6=-2\times 0.00632628,\quad
a_8=-2\times 0.0030085\quad \left( S=\frac 12\right) ;
$$
$$
a_2=-0.571428,\quad a_4=-0.0504597,\quad a_6=-0.0144762,\quad
a_8=-0.00656238 \quad \left( S=1\right) . 
$$
In the case of spin--$1$, the values of the local multipliers $m_k(s)$ are
increasing with increasing index $k$. In order to choose the starting term
in (110), we compared the values of $m_1(\infty )=1-a_2/4=1.143$ with
$m_2(\infty)=1+a_4/a_2=1.088$. I.e., an approximation cascade will have a 
more stable beginning if it starts from the second term in (110). Thus, we
come to the exponential approximants
\begin{equation}
\label{111}2E_2^{*}=-4+a_2x^2\exp \left( \frac{a_4}{a_2}x^2\tau \right) , 
\end{equation}
\begin{equation}
\label{112}2E_3^{*}=-4+a_2x^2\exp \left[ \frac{a_4}{a_2}x^2\exp \left( \frac{%
a_6}{a_4}x^2\tau \right) \right] . 
\end{equation}
From the minimal--difference condition
$\min _\tau \left| E_3^{*}(\tau )-E_2^{*}(\tau )\right|$
we find $\tau =1.568$ and $E_3^{*}(\tau ,S=1)=-2.328$, close to the result of
Pad\'e--summation, $E=-2.327\pm 0.001$ [76].

In the case of spin--$1/2$, the same approach is applicable, and we obtain 
$\tau =0.457$ and $4E_3^{*}(\tau,S=\frac12)=-2.665$. Pad\'e--summation gives 
in this case, $4E=-2.6785\pm 0.001$ [75].

Pass now to considering magnetization $M$. In the case of spin--$\frac 12$ 
it is more convenient to present $M$ in the form of a series in the parameter
$\delta$, related to $x$ by the equation  
\begin{equation}
\label{113}1-\delta =\left( 1-x^2\right) ^{1/2} .
\end{equation}
Then [75],
\begin{equation}
\label{114}2M \simeq 1+a_1\delta +a_2\delta ^2+a_3\delta ^3+a_4\delta 
^4+a_5\delta ^5 \qquad (\delta\rightarrow 0) ,
\end{equation}
$$
a_1=-\frac 49,\qquad a_2=0.08,\qquad a_3=-0.009319,\qquad a_4=-0.4642,\qquad
a_5=0.08257. 
$$
The local multipliers $m_k$ behave quite irregularly, reflecting the behavior 
of the coefficients. In this situation we resort to an aposteriori analysis of
the sequences of exponential approximants and corresponding aposteriori
multipliers, whose values will be given in brackets. For the sequence
including the constant term we observe a recurrent behavior, signaling an
emergence of a limiting cycle: 
\begin{equation}
\label{115}2M_1^{*}=\exp \left( a_1\delta \right) =0.641\qquad 
(m_1^{*}=1),
\end{equation}
\begin{equation}
\label{116}2M_2^{*}=\exp \left[ a_1\delta \exp \left( \frac{a_2}{a_1}\delta
\right) \right] =0.69 \qquad (m_2^{*}=0.737), 
\end{equation}
\begin{equation}
\label{117}2M_3^{*}=\exp \left[ a_1\delta \exp \left( \frac{a_2}{a_1}\delta
\exp \left( \frac{a_3}{a_2}\delta \right) \right) \right]
=0.685 \qquad (m_3^{*}=0.781), 
\end{equation}
\begin{equation}
\label{118}2M_4^{*}=\exp \left[ a_1\delta \exp \left( \frac{a_2}{a_1}\delta
\exp \left( \frac{a_3}{a_2}\delta \exp \left( \frac{a_4}{a_3}\delta \right)
\right) \right) \right] =0.641 \qquad (m_4^{*}=1), 
\end{equation}
\begin{equation}
\label{119}2M_5^{*}=\exp \left[ a_1\delta \exp \left( \frac{a_2}{a_1}\delta
\exp \left( \frac{a_3}{a_2}\delta \exp \left( \frac{a_4}{a_3}\delta \exp ( 
\frac{a_5}{a_4}\delta )\right) \right) \right) \right] =0.679\quad
(m_5^{*}=0.808). 
\end{equation}
We construct also a different sequence of exponential approximants, not
including into the renormalization procedure the constant term: 
\begin{equation}
\label{120}2M_2^{*}=1+a_1\delta \exp \left( \frac{a_2}{a_1}\delta \right)
=0.629 \qquad (m_1^{*}=1), 
\end{equation}
\begin{equation}
\label{121}2M_3^{*}=1+a_1\delta \exp \left[ \frac{a_2}{a_1}\delta \exp
\left( \frac{a_3}{a_2}\delta \right) \right] =0.621 \qquad (m_2^{*}=0.802), 
\end{equation}
\begin{equation}
\label{122}2M_4^{*}=1+a_1\delta \exp \left[ \frac{a_2}{a_1}\delta \exp
\left( \frac{a_3}{a_2}\delta \exp \left( \frac{a_4}{a_3}\delta \right)
\right) \right] =0.556 \quad (m_3^{*}=5.149\times 10^{-6}), 
\end{equation}
\begin{equation}
\label{123}2M_5^{*}=1+a_1\delta \exp \left( \frac{a_2}{a_1}\delta \exp
\left( \frac{a_3}{a_2}\delta \exp \left( \frac{a_4}{a_3}\delta \exp (\frac{%
a_5}{a_4}\delta )\right) \right) \right) =0.613 \quad (m_4^{*}=0.722). 
\end{equation}
The last value agrees well with the result of Ref.[75], $2M=0.605\pm 0.015$.

In the case of spin--$1,$ the following expansion in powers of $\delta $ can
be obtained [76]: 
\begin{equation}
\label{124}M \simeq 1+a_2\delta ^2+a_3\delta ^3+a_4\delta ^4+a_5\delta ^5
\qquad (\delta\rightarrow 0) ,
\end{equation}
$$
a_2=-0.326528,\quad a_3=0.326528,\quad a_4=-0.73216,\quad a_5=1.3001056. 
$$
The values of local multipliers $m_k$, in this case, suggest that inclusion 
into consideration of the last term destabilizes the trajectory.
In this situation we again resort to the aposteriori analysis. The 
following sequence of exponential approximants can be readily written 
down, with the corresponding multipliers, calculated at $\tau =1$, 
shown in brackets: 
\begin{equation}
\label{125}M_2^{*}=\exp \left( a_2\delta ^2\tau \right) =0.721\qquad 
(m_1^{*}=1), 
\end{equation}
\begin{equation}
\label{126}M_3^{*}=\exp \left[ a_2\delta ^2\exp \left( \frac{a_3}{a_2}\delta
\tau \right) \right] =0.887\qquad (m_2^{*}=0.206)
\end{equation}
\begin{equation}
\label{127}M_4^{*}=\exp \left[ a_2\delta ^2\exp \left( \frac{a_3}{a_2}\delta
\exp \left( \frac{a_4}{a_3}\delta \tau \right) \right) \right]
=0.746\qquad (m_3^{*}=0.931),
\end{equation}
\begin{equation}
\label{128}M_5^{*}=\exp \left[ a_2\delta ^2\exp \left( \frac{a_3}{a_2}\delta
\exp \left( \frac{a_4}{a_3}\delta \exp \left( \frac{a_5}{a_4}\delta \tau
\right) \right) \right) \right] =0.848\quad (m_4^{*}=0.33).
\end{equation}
We observe two subsequences, with odd and even numbers, probably
embracing the correct result from below and above, respectively. We can
suspect that they both define a stable quasifixed point, corresponding to
a focus. In order to locate it with the maximal possible precision, we 
impose the minimal difference condition on two points, belonging to two
different subsequences with the smallest values of multipliers, i.e.,
$\min _\tau \left| M_4^{*}(\tau )-M_3^{*}(\tau )\right|$. From this 
condition we determine $\tau =0.404$ and $M_4^{*}(\tau,S=1)=0.804,$ 
in agreement with the estimate $M=0.81\pm 0.01$ [76].

The expansions for the frequency moments, $\rho _1$ and $\rho _2$, of the
intensity of light scattering on the spin--pair excitations, are 
available for $S=1$ [76],
\begin{equation}
\label{129}2\rho_1 \simeq a_0+a_2x^2+a_3x^3+a_4x^4+a_5x^5 \qquad
(x\rightarrow 0) ,
\end{equation}
$$
a_0=14,\quad a_2=0.530612,\quad a_3=-0.141138,\quad a_4=0.033837,\quad
a_5=0.0171678; 
$$
\begin{equation}
\label{130}4\rho _2\simeq a_0+a_2x^2+a_3x^3+a_4x^4+a_5x^5 \qquad
(x\rightarrow 0) ,
\end{equation}
$$
a_0=196,\quad a_2=17.469,\quad a_3=-5.33453,\quad a_4=3.32992,\quad
a_5=0.626767. 
$$

Consider the case of $\rho _1$. Comparing the local multipliers, as 
$s\rightarrow \infty,\; m_1(\infty )=1+a_2/a_0=1.038$ and $m_2(\infty
)=1+a_3/a_2=0.734,$ we conclude that the constant term should not be included
into the renormalization procedure. Then, at $\tau =1$ and $x=1,$ the 
following values of the exponential approximants can be calculated: 
\begin{equation}
\label{131}2\left( \rho _1^{*}\right) _2=a_0+a_2x^2\exp \left( 
\frac{a_3}{a_2}x\tau \right) =7.203\qquad (m_1^{*}=1), 
\end{equation}
\begin{equation}
\label{132}2\left( \rho _1^{*}\right) _3=a_0+a_2x^2\exp \left( 
\frac{a_3}{a_2} x\exp \left( \frac{a_4}{a_3}x\tau \right) \right) =7.215 
\qquad (m_2^{*}=0.833), 
\end{equation}
\begin{equation}
\label{133}2\left( \rho _1^{*}\right) _4=a_0+a_2x^2\exp \left( 
\frac{a_3}{a_2} x\exp \left( \frac{a_4}{a_3}x\exp \left (
\frac{a_5}{a_4}x\tau \right )
\right ) \right ) =7.222 \quad (m_3^{*}=0.585). 
\end{equation}
We observe a smoothly behaving sequence of multipliers. From the minimal
difference condition
$\min _\tau \left| \left( \rho _1^{*}\right) _4-\left( \rho _1^{*}\right)
_3\right|$,
we find $\tau =1.59325$ and $\left( \rho _1^{*}\right) _4=7.221,$ in
agreement with $\rho _1=7.22\pm 0.02$, quoted in Ref. [76]. Identical
analysis leads to the value $\left( \rho _2^{*}\right) _4=52.804\ \left(
\tau =1.27\right)$, again in agreement with $\rho _2=53.0\pm 0.3$ from
Ref.[76]. The value of the parameter $R=(\sqrt{\rho _2-\rho _1^2})/\rho _1$
is equal to $0.113$, close to $0.12\pm 0.3~\ $from [76].

\section{Critical Phenomena}

\subsection{Martinelli--Parisi $\epsilon-$Expansion}

Martinelli and Parisi suggested an interesting way to control the 
position--space renormalization group calculations [77], connecting 
the approximate Migdal--Kadanoff transformation with the exact theory by 
means of the control shift parameter $\epsilon$, equal to zero for
Migdal--Kadanoff approximation and equal to one for the exact renormalization
transformation. This approach generates the expansions in powers of
$\epsilon$, considered as a small parameter, around the Migdal--Kadanoff
results. Finally, in order to reach the "exact" solution, one should set
$\epsilon =1$. The results can be further improved by imposing the
condition on zero derivative of physical quantities at $\epsilon =1$ [77,78].

The following expansion for the critical index $\nu$ of the Ising model 
is available [77] for the square lattice: 
\begin{equation}
\label{134}\nu ^{-1}\simeq w_0+w_1\epsilon +w_2\epsilon ^2 \qquad
(\epsilon\rightarrow 0) ,
\end{equation}
$$
w_0=0.687,\ w_1=1.14,\ w_2=-1.21. 
$$
Directly for the index $\nu$, one can find 
\begin{equation}
\label{135}\nu \simeq a_0+a_1\epsilon +a_2\epsilon ^2 \qquad
(\epsilon\rightarrow 0) ,
\end{equation}
$$
a_0=1.456,\quad a_1=-2.415,\quad a_2=6.572. 
$$
When only two starting terms from (135) are considered, the result is
definitely wrong, $\nu =-0.96$, and with three terms we get $\nu=5.61$. 
This shows how expansion (135) is bad. The renormalization procedure, not
including into consideration the constant term, gives the exponential
approximant
\begin{equation}
\label{136}\nu _2^{*}(\epsilon )=a_0+a_1\epsilon \exp \left( \frac{a_2}{a_1}%
\epsilon \right) =1.297.
\end{equation}
This result, is, probably, too large and it does change much, to the value 
$0.568$, when the condition on zero derivative is imposed. Then, in order to
extend the validity of the expansion (135), let us add to it a
negative trial term $-\left| a_3\right| \epsilon ^3$.
The following renormalized expression can be written: 
\begin{equation}
\label{137}
\nu _3^{*}(\epsilon ,a_3)=a_0+a_1\epsilon \exp \left( 
\frac{a_2}{a_1}\epsilon 
\frac 1{1+\frac{\left| a_3\right| }{a_2}\epsilon }\right) .
\end{equation}
From the condition equivalent to (45),
$$
\frac{\partial \nu _3^{*}(\epsilon ,a_3)}{\partial \epsilon }=0, 
$$
at $\epsilon =1$, we find $a_3=4.268$ and $\nu _3^{*}(\epsilon =1)=0.992$,
in excellent agreement with the exact result, $\nu =1.$ Our estimate is much
better than the result of Pad\'e--summation $\nu =0.945$, quoted in [77].

For the critical temperature $T_c$ the following expansion was obtained [77]: 
\begin{equation}
\label{138}T_c^{-1}\simeq b_0+b_1\epsilon +b_2\epsilon ^2 \qquad
(\epsilon\rightarrow 0) ,
\end{equation}
$$
b_0=0.4359,\qquad b_1=0.024,\qquad b_2=-0.109 .
$$
This gives for $T_c$
\begin{equation}
\label{139}T_c \simeq a_0+a_1\epsilon +a_2\epsilon ^2 \qquad
(\epsilon\rightarrow 0) ,
\end{equation}
$$
a_0=2.294,\quad a_1=-0.126,\quad a_2=0.581. 
$$
Two exponential approximants can be written,
\begin{equation}
\label{140}\left( T_c^{*}\right) _1=a_0\exp \left( \frac{a_1}{a_0}\epsilon
\tau \right) , 
\end{equation}
\begin{equation}
\label{141}\left( T_c^{*}\right) _2=a_0\exp \left[ \frac{a_1}{a_0}\epsilon
\exp \left( \frac{a_2}{a_1}\epsilon \tau \right) \right] . 
\end{equation}
From the minimal difference condition, at $\epsilon =1,$ we find $\tau
=0.278$ and $\left( T_c^{*}\right) _2=2.259$. The result does not change much
(to the value $2.248,\ \tau =0.217$) when the condition on zero derivative
of $\left( T_c^{*}\right) _2$ is imposed. Let us add to the expansion (139)
one more term $-\left| a_3\right| \epsilon ^3,$ so that the
following approximant can be written: 
\begin{equation}
\label{142}\left( T_c^{*}\right) _3=a_0\exp \left[ \frac{a_1}{a_0}\epsilon
\exp \left( \frac{a_2}{a_1}\epsilon \frac 1{1+\frac{\left| a_3\right| }{a_2}%
\epsilon }\right) \right] , 
\end{equation}
an determine $a_3$ from the condition on zero derivative at $\epsilon =1.$
Then, $\left| a_3\right| =0.664$ and $\left( T_c^{*}\right) _3=2.279$. Pad\'e
approximants in this case give a close result, $T_c=2.275$ [77].
This is to be compared with the exact $T_c=2.269$.

On a triangular lattice, the following expression for $\nu $ was obtained
[78] : 
\begin{equation}
\label{143}2^{1/\nu }\simeq a_0+a_1\epsilon +a_2\epsilon ^2 \qquad
(\epsilon\rightarrow 0) ,
\end{equation}
$$
a_0=1.6786,\qquad a_1=0.5344,\qquad a_2=-0.3952. 
$$
In this case, the exponential approximant 
\begin{equation}
\label{144}\left( 2^{1/\nu }\right) ^{*}=a_0\exp \left[ \frac{a_1}{a_0}%
\epsilon \exp \left( \frac{a_2}{a_1}\epsilon \right) \right] ,
\end{equation}
leads to $\nu ^{*}=1.035$, which is a much better value than $1.161$ 
obtained in Ref.[78] directly from (143). It is possible to improve our 
estimate performing the last step of the self--similar bootstrap along 
the most stable available trajectory, with the stabilizer $s$ corresponding 
to zero value of the local multiplier $m=1+a_2(1+s)/a_1s$. This yields 
\begin{equation}
\label{145}\left( 2^{1/\nu }\right) ^{*}=a_0\exp \left[ \frac{a_1}{a_0}
\epsilon \left( \frac s{s-\frac{a_2}{a_1}\epsilon \tau }\right) ^s\right]
,\qquad \ s=-\frac{a_2\epsilon }{a_1+a_2\epsilon }=2.86.
\end{equation}
Then, $\nu ^{*}=1.015$ for $\tau=1$. The derivative of $\nu ^{*}$ is equal 
to $0.035$. At the point $\tau =1.152$, the derivative goes to zero, and 
our estimate changes slightly to $\nu ^{*}=1.036.$

For the inverse critical temperature on a triangular lattice [78], we have
\begin{equation}
\label{146}T_c^{-1}\simeq b_0+b_1\epsilon +b_2\epsilon ^2 \qquad
(\epsilon\rightarrow 0) ,
\end{equation}
$$
b_0=0.3047, \qquad b_1=-0.0976, \qquad b_2=0.0501. 
$$
For the critical temperature we obtain 
\begin{equation}
\label{147}T_c\simeq a_0+a_1\epsilon +a_2\epsilon ^2 \qquad
(\epsilon\rightarrow 0) ,
\end{equation}
$$
a_0=3.282,\qquad a_1=1.051,\qquad a_2=-0.203. 
$$
The following exponential approximant is favored from the viewpoint of local
multipliers: 
\begin{equation}
\label{148}T_c^{*}=a_0+a_1\epsilon \exp \left( \frac{a_2}{a_1}\epsilon \tau
\right) .
\end{equation}
The derivative of (148) is quite large at $\tau =1$, so we
resort to the condition 
$$
\frac{\partial T_c^{*}}{\partial \epsilon }=0, 
$$
and at $\epsilon =1$ we find $\tau =-a_2/a_1=5.177$, and 
$$
T_c^{*}=a_0+a_1e^{-1}=3.669, 
$$
deviating from the exact value $3.642$ with the percentage error $0.741\%$,
while the result $3.888$, following from expansion (146) and quoted in [78], 
gives the error $6.755\%$.

For the first coefficient of the beta function $\beta _1$ of the
two--dimensional non--linear sigma--model the following expansion was 
obtained [79]: 
\begin{equation}
\label{149}\beta _1\simeq \frac{\sqrt{3}}{\ln (2)}\left( a_0+a_1\epsilon
+a_2\epsilon ^2\right) \qquad (\epsilon\rightarrow 0) ,
\end{equation}
$$
a_0=-\frac 1{24},\qquad a_2=\frac 3{64},\qquad a_3=-\frac{69}{86}. 
$$
The exponential approximant 
\begin{equation}
\label{150}\beta _1^{*}=\frac{\sqrt{3}}{\ln (2)}\left( a_0+a_1\epsilon \exp
\left( \frac{a_2}{a_1}\epsilon \right) \right) =-0.081
\end{equation}
is in good agreement with the exact result $\beta_1=-\frac 1{4\pi }=-0.08$ 
[79]. The derivative of $\beta _1^{*}$ is small and equals to $0.006$, so 
that we can safely stop at this point.

\subsection{Localization Length}

The critical exponent $\nu$ describing the divergence of the localization
length in the vicinity of the Anderson transition from the insulating to the
conducting phase, 
$$
l\sim (E_c-E)^{-\nu }, 
$$
as the energy $E$ of an electron approaches the mobility edge $E_c,$ can be
presented in the form of $(2+\epsilon )$--expansion, or $(d-2)$--expansion,
where $d$ is the dimensionality of space [80], 
\begin{equation}
\label{151}\nu \simeq \frac 1\epsilon +b\epsilon ^2+c\epsilon ^3 \qquad
(\epsilon\rightarrow 0) ,
\end{equation}
$$
b=-\frac 94\zeta (3),\quad c=\frac{27}{16}\zeta (4). 
$$
The result given by the starting two terms is wrong, $\nu =-1.705,$ while it
is known that $\nu \geq 2/3$ in the three--dimensional case ($\epsilon =1)$
[81]. The third term slightly improves the situation, but the result remains
small, $\nu =0.122$. Renormalizing the starting two terms yields the
exponential approximant 
\begin{equation}
\label{152}\nu ^{*}\left( \epsilon ,\tau \right) =\frac 1\epsilon \exp
\left( b\epsilon ^3\tau \right) . 
\end{equation}
Let us impose the condition 
$$
\frac{\partial\nu^*\left (\epsilon,\tau\right )}{\partial\epsilon} =0, 
$$
discussed above in Section IV. Then, the control function $\tau (\epsilon )$
is found as follows: 
\begin{equation}
\label{153}\tau (\epsilon )=\frac 1{3b\epsilon^3}. 
\end{equation}
At $\epsilon =1$, we obtain $\tau =-0.123246$, corresponding to the necessity
to move backwards, from the wrong point of the approximation cascade.
Our estimate for the critical index $\nu ^{*}=\exp
\left( \frac 13\right) =1.39561$ better agrees with the numerical result 
$1.3$ [82] than the Pad\'e--Borel estimate $0.730$ quoted in Ref.[80].

\subsection{Amplitude Ratios}

Consider the $3d$ Ising model. Different amplitude ratios are
available in the form of the Wilson $\epsilon$--expansion $(\epsilon=4-d)$
around the dimensionality four [83]. For the ratio $C^{+}/C^{-}$, related 
to the magnetic susceptibility in zero field, the following expression is
available: 
\begin{equation}
\label{154}\frac{C^{+}}{C^{-}}\simeq 2^\gamma \left[ 1+a_1\epsilon 
+a_2\epsilon ^2+a_3\epsilon ^3\right] 
\qquad (\epsilon\rightarrow 0) , 
\end{equation}
$$
a_1=\frac 12,\qquad a_2=\frac{25}{108},\qquad a_3=\frac 1{24}\lambda +\frac
1{36}\zeta (3)+\frac{1159}{11664}\ ,\qquad \lambda =1.171953.
$$
In order to improve the stability of the procedure, let us invert 
$C^{+}/C^{-}$ and study 
\begin{equation}
\label{155}\left( \frac{C^{+}}{C^{-}}\right) ^{-1}\simeq 2^{-\gamma }\left[
1+b_1\epsilon +b_2\epsilon ^2+b_3\epsilon ^3\right] 
\qquad (\epsilon\rightarrow 0) , 
\end{equation}
$$
b_1=-\frac 12,\qquad b_2=\frac 1{54},\qquad b_3=-7.5519\ 10^{-2}. 
$$
From the viewpoint of stability conditions, the following two approximants
are well justified: 
\begin{equation}
\label{156}\left( \left( \frac{C^{+}}{C^{-}}\right) ^{-1}\right)
_2^{*}=2^{-\gamma }\left[ 1+b_1\epsilon \exp \left( \frac{b_2}{b_1}\epsilon
\tau \right) \right] , 
\end{equation}
\begin{equation}
\label{157}\left( \left( \frac{C^{+}}{C^{-}}\right) ^{-1}\right)
_3^{*}=2^{-\gamma }\left[ 1+b_1\epsilon \exp \left( \frac{b_2}{b_1}\epsilon
\exp (\frac{b_3}{b_2}\epsilon \tau )\right) \right] , 
\end{equation}
and from the minimal difference condition 
$$
\min _\tau \left| \left( \left( \frac{C^{+}}{C^{-}}\right) ^{-1}\right)
_3^{*}-\left( \left( \frac{C^{+}}{C^{-}}\right) ^{-1}\right) _2^{*}\right| , 
$$
with the typical value of $\gamma=1.24$,
we obtain $\tau =0.297$. Correspondingly, $\left( C^{+}/C^{-}\right)
^{*}=4.673$. This value agrees well with the theoretical estimates and
experimental data [83]. 

The ratio$\ A^{+}/A^{-}$ is related to the two-point correlation function
at zero momentum, 
\begin{equation}
\label{158}
A^{+}/A^{-}\simeq 2^{\alpha -2}\left[ 1+\epsilon -\left| a_2\right|
\epsilon ^2\right ]  \qquad (\epsilon\rightarrow 0) , 
\end{equation}
$$
a_2=\frac{43}{54}-\frac 16\lambda -\zeta (3)\ . 
$$
Two renormalized expressions can be constructed,
\begin{equation}
\label{159}
\left( \frac{A^{+}}{A^{-}}\right) _1^{*}=2^{\alpha -2}\left[ \exp
(\epsilon \tau )\right] , 
\end{equation}
\begin{equation}
\label{160}\left( \frac{A^{+}}{A^{-}}\right) _2^{*}=2^{\alpha -2}\left[ \exp
\left( \epsilon \exp \left( -\left| a_2\right| \epsilon \tau \right) \right)
\right] , 
\end{equation}
both exponential approximants being justified from the viewpoint of
stability conditions. From the condition
$$
\min _\tau \left| \left( \frac{A^{+}}{A^{-}}\right) _2^{*}-\left( 
\frac{A^{+} }{A^{-}}\right) _1^{*}\right| ,  
$$
with the typical value of $\alpha=0.11$, we find that $\tau =0.669$ and 
$\left( \frac{A^{+}}{A^{-}}\right)_2^{*}=0.527$, well agreeing with 
the data of Table 5 from Ref. [83]. 

For the amplitude ratios $R_c$ and $R_\chi$ (see [83]),
the following expansions are available: 
\begin{equation}
\label{161}R_c\simeq \frac 192^{-2\beta -1}\epsilon \left[ 1+a_1\epsilon 
-\left| a_2\right| \epsilon ^2\right] 
\qquad (\epsilon\rightarrow 0) , 
\end{equation}
$$
a_1=\frac{17}{27},\qquad  a_2=\frac{989}{2916}-\frac 49\zeta (3)-
\frac 23\lambda \ , 
$$
and 
\begin{equation}
\label{162}R_\chi \simeq 3^{(\delta -3)/2}2^{\gamma +(1-\delta )/2}\left[ 
1+\left(
\frac 1{72}+\frac 1{36}\lambda -\frac 1{18}\zeta (3)\right) \epsilon
^3\right] 
\qquad (\epsilon\rightarrow 0) .
\end{equation}
The self--similar exponential approximants are
\begin{equation}
\label{163}\left( R_c\right) _1^{*}=\frac 192^{-2\beta -1}\epsilon \left[
\exp \left( a_1\epsilon \tau \right) \right] ,
\end{equation}
\begin{equation}
\label{164}\left( R_c\right) _2^{*}=\frac 192^{-2\beta -1}\epsilon \left[
\exp \left( a_1\epsilon \exp \left( -\frac{\left| a_2\right| }{a_1}\epsilon
\tau \right) \right) \right] . 
\end{equation}
From the minimal difference condition 
$\min _\tau \left| \left( R_c\right) _2^{*}-\left( R_c\right) _1^{*}\right|$,
with the typical value $\beta=0.325$, we obtain $\tau =0.633$ and 
$\left( R_c\right) _2^{*}=0.053$ agreeing well with the results quoted in 
Ref.[83]. 

The renormalized expression for $R_{\chi}$ writes
\begin{equation}
\label{165}R_\chi ^{*}=3^{(\delta -3)/2}2^{\gamma +(1-\delta )/2}\left[ \exp
\left( \frac 1{72}+\frac 1{36}\lambda -\frac 1{18}\zeta (3)\right) \epsilon
^3\right] .
\end{equation}
We take $\delta =4.825$, corresponding, by virtue of the
scaling relation, to a reasonable value for the critical index $\eta =0.03$.
Then $R_\chi ^{*}=1.675$, close to various estimates presented in [83].

Finally, we consider the quantity $\left| z_0\right|$, the universal rescaled
spontaneous magnetization, 
\begin{equation}
\label{166}\left| z_0\right| \simeq \sqrt{3}2^\beta \left[ 1+a_1\epsilon
+a_2\epsilon ^2-\left| a_3\right| \epsilon ^3\right] 
\qquad (\epsilon\rightarrow 0) , 
\end{equation}
$$
a_1=\frac 14,\qquad a_2=\frac{73}{864},\qquad a_3=\frac 1{24}\lambda -\frac
7{36}\zeta (3)+\frac{5581}{93312}\ . 
$$
The self--similar exponential approximants, justified from the viewpoint
of an aposteriori stability analysis, are
\begin{equation}
\label{167}\left| z_0\right| _2^{*}=\sqrt{3}2^\beta \left[ \exp \left(
a_1\epsilon \exp \left( \frac{a_2}{a_1}\epsilon \tau \right) \right) \right] ,
\end{equation}
\begin{equation}
\label{168}\left| z_0\right| _3^{*}=\sqrt{3}2^\beta \left[ \exp \left(
a_1\epsilon \exp \left( \frac{a_2}{a_1}\epsilon \exp \left( -\frac{\left|
a_3\right| }{a_2}\epsilon \tau \right) \right) \right) \right] , 
\end{equation}
and from the minimal difference condition
$\min _\tau \left| \left| z_0\right| _3^{*}-\left| z_0\right| _2^{*}\right|$
we find $\tau =0.633$ and $\left| z_0\right| _3^{*}=2.913$ agreeing 
well with the result of Pad\'e--summation $2.87\pm 0.06$ [83].

\subsection{Critical Indices from the Wilson $\epsilon$--Expansion}

Critical indices are usually obtained from the Wilson $\epsilon -$expansion
[84,85] using some kind of a resummation procedure. As a rule, for
different values of $n$, standing for the number of the order parameter
components, one obtains some values not related to each other analytically.
We obtain below {\it analytical} renormalized expressions for the critical
indices $\nu $ and $\eta $, valid for arbitrary $n$.

For the critical index $\eta$, the expansion is available up to the fifth
order term in $\epsilon $ [85]. For convenience, we reproduce the
higher-order coefficients in Appendix, 
\begin{equation}
\label{169}\eta \simeq a_2(n)\epsilon ^2+a_3(n)\epsilon ^3+a_4(n)\epsilon
^4+a_5(n)\epsilon ^5
\qquad (\epsilon\rightarrow 0) , 
\end{equation}
$$
a_2(n)=\frac{n+2}{2(n+8)^2},\qquad a_3(n)=\frac{n+2}{8(n+8)^4}(272+56n-n^2),
\qquad \ldots
$$
The following exponential approximants are justified from the viewpoint of
the stability conditions for local multipliers: 
\begin{equation}
\label{170}\eta _4^{*}(\epsilon ,n,\tau )=a_2(n)\epsilon ^2+a_3(n)\epsilon
^3\exp \left( \frac{a_4(n)}{a_3(n)}\epsilon \tau \right) , 
\end{equation}
\begin{equation}
\label{171}\eta _5^{*}(\epsilon ,n,\tau )=a_2(n)\epsilon ^2+a_3(n)\epsilon
^3\exp \left[ \frac{a_4(n)}{a_3(n)}\epsilon \exp \left( \frac{a_5(n)}{a_4(n)}%
\epsilon \tau \right) \right] , 
\end{equation}
From the minimal difference condition 
$\min _\tau \left| \eta _5^{*}(\epsilon ,n,\tau )-
\eta _4^{*}(\epsilon,n,\tau )\right| , $
equivalent in this case to the equation
$$
\tau =\exp \left( \frac{a_5(n)}{a_4(n)}\epsilon \tau \right) , 
$$
we can find the control function $\tau =\tau (n)$ at $\epsilon =1$. 
From formula (171), we can estimate the critical index
$\eta _5^{*}(1,n,\tau(n))\equiv \eta ^{*}(n)$. The results of calculations 
are presented in the Table. They agree well with the majority of 
different kinds of estimates available for the critical index $\eta$.

For the critical index $\nu$, up to the fifth order in $\epsilon $ (see
Appendix), one has [85] 
\begin{equation}
\label{172}\nu ^{-1}\simeq b_0(n)+b_1(n)\epsilon +b_2(n)\epsilon 
^2+b_3(n)\epsilon ^3+b_4(n)\epsilon ^4+b_5(n)\epsilon ^5
\qquad (\epsilon\rightarrow 0) , 
\end{equation}
$$
b_0(n)=2,\qquad b_1(n)=-\frac{n+2}{n+8}, \qquad 
b_2(n)=-\frac{(n+2)(13n+44)}{2(n+8)^3}\; , \qquad \ldots 
$$
The following exponential approximants, preserving the correct limits $\nu
=\frac{1}{2}$ at $n=-2$ [86-88] and $\nu =1$, as $n\rightarrow \infty$ [89], 
can be written, giving at $\tau =1$ and $\epsilon =1$ the following results
\begin{equation}
\label{173}\left( \nu ^{-1}\right) _2^{*}=b_0(n)+b_1(n)\epsilon \exp \left[ 
\frac{b_2(n)}{b_1(n)}\epsilon \tau \right] , 
\end{equation}
$$
\nu _2^{*}(n=0)=0.607,\quad \nu _2^{*}(n=1)=0.655,\quad \nu
_2^{*}(n=2)=0.698,\quad \nu _2^{*}(n=3)=0.736. 
$$
\begin{equation}
\label{174}\left( \nu ^{-1}\right) _3^{*}=b_0(n)+b_1(n)\epsilon \exp \left[ 
\frac{b_2(n)}{b_1(n)}\epsilon \exp \left( \frac{b_3(n)}{b_2(n)}\epsilon \tau
\right) \right] , 
\end{equation}
$$
\nu _3^{*}(n=0)=0.579,\quad \nu _3^{*}(n=1)=0.616,\quad \nu
_3^{*}(n=2)=0.651,\quad \nu _3^{*}(n=3)=0.683. 
$$
\begin{equation}
\label{175}\left( \nu ^{-1}\right) _4^{*}=b_0(n)+b_1(n)\epsilon \exp \left[ 
\frac{b_2(n)}{b_1(n)}\epsilon \exp \left( \frac{b_3(n)}{b_2(n)}\epsilon \exp
\left( \frac{b_4(n)}{b_3(n)}\epsilon \tau \right) \right) \right] , 
\end{equation}
$$
\nu _4^{*}(n=0)=0.603,\quad \nu _4^{*}(n=1)=0.649,\quad \nu
_4^{*}(n=2)=0.692,\quad \nu _4^{*}(n=3)=0.729. 
$$
\begin{equation}
\label{176}\left( \nu ^{-1}\right) _5^{*}=b_0(n)+b_1(n)\epsilon \exp \left[ 
\frac{b_2(n)}{b_1(n)}\epsilon \exp \left( \frac{b_3(n)}{b_2(n)}\epsilon \exp
\left( \frac{b_4(n)}{b_3(n)}\epsilon \exp \left( \frac{b_5(n)}{b_4(n)}
\epsilon \tau \right) \right) \right) \right] . 
\end{equation}
$$
\nu _5^{*}(n=0)=0.58,\quad \nu _5^{*}(n=1)=0.618,\quad \nu
_5^{*}(n=2)=0.654,\quad \nu _5^{*}(n=3)=0.688. 
$$
Two sequences, with odd and even numbers, are clearly seen. The last two
approximants give the closest values. From the minimal difference condition 
$\min _\tau \left| \left( \nu ^{-1}\right) _5^{*}-
\left( \nu ^{-1}\right)_4^{*}\right|$, which simply reduces to the equation 
$$
\tau =\exp \left( \frac{b_5(n)}{b_4(n)}\epsilon \tau \right) , 
$$
one can easily find the control function $\tau =\tau (n),$ and, finally,
calculate the critical index, $\nu ^{*}(n)=\left( \left( \nu ^{-1}\right)
_5^{*}\right) ^{-1}.$

For the critical exponent $\omega (\epsilon )=2B_\epsilon ^{^{\prime }}$,
calculated in the renormalized infrared-stable fixed point, the following
expansion is available [85]: 
\begin{equation}
\label{177}\omega (\epsilon )\simeq \epsilon +
c_2(n)\epsilon ^2+c_3(n)\epsilon ^3
\qquad (\epsilon\rightarrow 0) , 
\end{equation}
$$
c_2(n)=-\frac{9n+42}{\left( n+8\right) ^2}\; ,
$$
$$
c_3(n)=\frac
1{4(n+8)^4}\left[ 33n^3+538n^2+4288n+9658+\zeta (3)\left( n+8\right)
96(5n+22)\right] . 
$$
We limit here the discussion by the third--order terms, since the higher
order terms grow impetuously, jeopardizing the fulfillment of the stability
conditions. The following exponential approximants are available being
well justified from the viewpoint of stability,
\begin{equation}
\label{178}\omega _2^{*}(\epsilon ,n,\tau )=\epsilon \exp \left(
c_2(n)\epsilon \tau \right) 
\end{equation}
\begin{equation}
\label{179}\omega _3^{*}(\epsilon ,n,\tau )=\epsilon \exp \left[
c_2(n)\epsilon \exp \left( \frac{c_3(n)}{c_2(n)}\epsilon \tau \right)
\right] , 
\end{equation}
and from the minimal difference condition 
$$
\tau =\exp \left( \frac{c_3(n)}{c_2(n)}\epsilon \tau \right) , 
$$
one can find the control function $\tau =\tau (n)$ at $\epsilon =1$. We
obtain the following values of the index in the physically interesting
region:%
$$
\omega ^{*}(n=0)=0.788,\quad \omega ^{*}(n=1)=0.788,\quad \omega
^{*}(n=2)=0.791,\quad \omega ^{*}(n=3)=0.794. 
$$
The limiting values, $\omega ^{*}(n=-2)\approx 0.8,\; \omega^{*}(n
\rightarrow \infty )\approx 1$, sound reasonable. The dynamical critical 
index $\theta =\omega \nu$ can be estimated using (176) and (179). The 
results for the critical indices $\eta$ and $\nu $, presented in the
Table, well agree with other theoretical estimates and available experimental 
data [90-92]. Let us stress again, that at $n=-2$ and $n\rightarrow 
\infty$, we obtain the exact results.

\subsection{Critical Indices from the Field Theory Expansion}

Field--theory approach in the theory of critical phenomena is, usually,
very accurate [92,93]. In Ref [13] we analyzed the expansions in powers of
the interaction constant $g$ ($g$--expansion) for some critical indices from
the viewpoint of the limiting cases $n=-2,\; n\rightarrow\infty$, and found
by a direct inspection of the expressions for $\eta $ and $\gamma $ from
[92], that the $n\rightarrow\infty$ limit, corresponding to the spherical 
model [89], is obeyed rigorously if $g=1$, i.e. $\eta =0,\; \gamma =2$, and
the $n=-2$ limit, corresponding to the gaussian polymer [86-88], is obeyed
with a very high accuracy for arbitrary $g$, i.e. $\eta\approx 0,\;\gamma
\approx 1$.

The standard approach [93] uses, for computing the renormalized infrared
stable fixed point $g^{*}$ of the beta--function $B_1(g)$, a complicated
Borel summation technique. Then critical indices are calculated as $\gamma
(g^{*})$ and $\eta(g^{*})$. The results, thus, depend on the way in which the
position of the fixed point is determined, although, different approaches
give the results very close to each other [92]. We suggest below a simple
way to minimize an uncertainty related to the position of the fixed point.
Let us use below two different approaches to the determination of $g^{*}$.
The first approach was suggested in Ref.[13]. It is based on three 
starting terms from the $g$--expansion from Ref.[92] (see Appendix): 
\begin{equation}
\label{180}B_1(g)\simeq -g+g^2+a_3(n)g^3+a_4(n)g^4+a_5(n)g^5+a_6(n)g^6
\qquad (g\rightarrow 0) ,
\end{equation}
$$
a_3(n)=-\frac{6.07407408\ n+28.14814815}{(n+8)^2}\; ,\qquad \ldots
$$
For the velocity field we have 
$$ v_3(f)=-\frac{a_3}8\left(\frac 12(1+\sqrt{1+4f}\right) ^3 , $$   
which is to be substituted into the evolution integral 
\begin{equation}
\label{181}\int_{-g+g^2}^{B_1^{*}}\frac{df}{\text{{\it v}}_3(f)}=\tau .
\end{equation}
The root $\ g_1^{*}=g_1^{*}(n,\tau )$ of the equation $\ B_1^{*}(g,n,\tau
=1)=0$ \ is obtained numerically, as a function $\ g_1^{*}=g_1^{*}(n).$ The
following values were obtained in the physically important cases:
$$
g_1^{*}(n=0)=1.59,\quad g_1^{*}(n=1)=1.559,\quad g_1^{*}(n=2)=1.524,\quad
g_1^{*}(n=3)=1.491. 
$$
Such a decreasing, with $n$, dependence is characteristic to the  majority 
of related studies. We observed also, that at $n=-2,\; g_1^{*}=1.599$ 
and at $n\rightarrow\infty,\; g_1^{*}=1$. The dependence of $g^{*}(n)$ in the
interval $n\in (-2,0)$ is nonmonotonous, a maximum is reached at $n=-1$, 
where $g_1^{*}=1.61$. We also constructed a different beta--function 
$B_2(g)$,
\begin{equation}
\label{182}B_2(g,\tau )=-g\exp \left( -a_3(n)g^2\right) +g^2\exp \left(
a_4(n)g^2\right) +a_5(n)g^5\exp \left( \frac{a_6(n)}{a_5(n)}g\tau \right) , 
\end{equation}
leading to the following increasing with $n$ values of the fixed point 
$g_2^{*}(n,\tau)$ at $\tau =1$,
$$
g_2^{*}(n=0)=1.163,\quad g_2^{*}(n=1)=1.208,\quad g_2^{*}(n=2)=1.249,\quad
g_2^{*}(n=3)=1.285, 
$$
possessing a maximum at $n=7$. Such an increasing dependence better 
corresponds to the known decrease of the critical temperature $T_c\;$, from
the Ising $(n=1)$ to Heisenberg $(n=3)$ model [72], since $T_c\sim g^{-1}$ 
[94]. We attempt now to minimize uncertainty, connected to the way of 
determining the fixed point. Imposing the minimal difference condition 
$$
\min _\tau\left| g_2^{*}(n,\tau )-g_1^{*}(n,\tau )\right| , 
$$
we obtain the control function $\tau =\tau (n)$, and the following values
for the optimized zero of the beta--function
$$
g^{*}(n=0)=1.311,\quad g^{*}(n=1)=1.334,\quad g^{*}(n=2)=1.352,\quad
g^{*}(n=3)=1.365 .
$$
These values increase till $n=5$, when $g^{*}(n=5)=1.372$, and then 
decrease till $g^{*}=1$, as $n\rightarrow \infty$.

For the critical index $\eta ,$ we keep all the terms available$,$ up to the
sixth order in powers of $g$ [92] (see Appendix): 
\begin{equation}
\label{183}\eta(g)\simeq b_2(n)g^2+b_3(n)g^3+b_4(n)g^4+b_5(n)g^5+b_6(n)g^5
\qquad (g\rightarrow 0) ,
\end{equation}
$$
b_2(n)=\frac{0.2962962963(n+2)}{(n+8)^2}\; ,\quad b_3(n)=\frac{
0.0246840014n^2+0.246840014n+0.3949440224}{(n+8)^3}\; , \ldots 
$$

From the aposteriori stability analysis we select the fifth order 
approximant corresponding to the smallest multiplier:
\begin{equation}
\label{186}\eta _5^{*}(g,n)=b_2(n)g^2\exp \left[ \frac{b_3(n)}{b_2(n)}g\exp
\left( \frac{b_4(n)}{b_3(n)}g\exp \left( \frac{b_5(n)}{b_4(n)}g\right)
\right) \right] , 
\end{equation}
so that 
$$
\eta _5^{*}(g^{*},n=0)=0.027,\quad \eta _5^{*}(g^{*},n=1)=0.034,\quad \eta
_5^{*}(g^{*},n=2)=0.039,\quad \eta _5^{*}(g^{*},n=3)=0.04. 
$$
These values agree well with those quoted in Refs. [91,92].

For the critical index $\gamma $ we write down all the terms available, in
powers of $g$ [92] (see Appendix), 
\begin{equation}
\label{188}\gamma^{-1}\simeq
1+c_1(n)g+c_2(n)g^2+c_3(n)g^3+c_4(n)g^4+c_5(n)g^5+c_6(n)g^6
\qquad (g\rightarrow 0) ,
\end{equation}
$$
c_1(n)=-\frac{n+2}{2(n+8)},\qquad c_2(n)=\frac{n+2}{(n+8)^2}, \qquad \ldots
$$
The third order exponential approximant is selected by an aposteriori
analysis, since it corresponds to the smallest value of the multiplier, 
giving the following expression for the critical index:
\begin{equation}
\label{190}\left( \gamma ^{-1}\right) _3^{*}=1+c_1(n)g\exp \left[ \frac{%
c_2(n)}{c_1(n)}g\exp \left( \frac{c_3(n)}{c_2(n)}g\right) \right] , 
\end{equation}
so that
$$
\gamma ^{*}(g^{*},n=0)=1.164,\quad \gamma ^{*}(g^{*},n=1)=1.243,\quad \gamma
^{*}(g^{*},n=2)=1.319,\quad \gamma ^{*}(g^{*},n=3)=1.39,\quad 
$$
which again agrees well with the data of Refs.[90-92].

\section{Discussion}

In this paper we have developed an {\it analytical} approach for summing 
divergent series with arbitrary noninteger as well as integer powers. This
approach is based on the novel notion of the {\it self--similar exponential 
approximants}. By a number of examples we show that the developed method
is general and accurate. In addition, the exponential approximants have a
simple analytical structure, even for quite large number of perturbative
terms used, when one usually has to resort to numerical techniques. Because 
of its analytical nature, our method permits one to accomplish direct 
analysis of resulting formulas with respect to the variation of physical
parameters. All examples we have analyzed are related to important physical 
phenomena, such as critical and crossover phenomena. We demonstrate that
the method of self--similar exponential approximants provides an effective
general tool for treating many different problems of statistical physics.

In conclusion, it is worth touching the following question. Assume that we
are given a truncated series (1). Then, would it be possible, looking at
the given series, to get some heuristic arguments, when the self--similar
approximants in the form of nested exponentials should yield good results.
The answer to this question is: Yes, we can made a preliminary estimates
whether the nested exponentials would work well. Such a first--glance
investigation can be done by analysing the apriori multiplier (42). Since
the case of the nested exponentials corresponds to $s\rightarrow\infty$,
then from (48), we have
\begin{equation}
\lim_{s\rightarrow\infty} m_k(x,s) = 
\sum_{n=0}^k\frac{a_n}{a_0}x^{\alpha_n -\alpha_0} .
\end{equation}
From here, we immediately notice what would be the favourable cases for
the better stability of the procedure, which is related to the condition
of the minimal multiplier modulus, $|m_k(x,\infty)|$. Such favorable cases
include: (i) When $a_n$ decreases as $n$ increases, so that $|a_n/a_0|<1$
and $a_n/a_0\rightarrow 0$, as $n\rightarrow\infty$. If 
$|x^{\alpha_n-\alpha_0}|$ increases with $n$, then, to compensate this
increase, $|a_n/a_0|$ must decrease sufficiently fast. (ii) When with
increasing $n$, $|x^{\alpha_n-\alpha_0}|$ decreases. This kind of
situation occurs, e.g., in the strong--coupling limit of many quantum
problems, when $\alpha_n<0$, and $x^{\alpha_n-\alpha_0}\rightarrow 0$, as
$n\rightarrow\infty$. If the decrease of $|x^{\alpha_n-\alpha_0}|$ is
sufficiently fast, then $|a_n/a_0|$ may even grow with $n$. (iii) When (1)
is an alternating series, that is, the coefficients $a_n$ change their
signs with changing $n$. In this case, even if neither $|a_n/a_0|$ nor
$|x^{\alpha_n-\alpha_0}|$ decrease, but nevertheless, because of
alternating signs of $a_n$, the value $|m_k(x,\infty)|$ may be small.

The worst case, as is seen from (187), would be when all coefficients
$a_n$ are of the same sign, $a_n$ increases with $n$, so that $a_n/a_0>1$,
and in addition, when $x^{\alpha_n-\alpha_0} > 1$. To explain why this
case is really the worst, consider a simple illustration for series (1)
truncated at second order,
$$ p_2(x) = a_0 + a_1x + a_2x^2 . $$
The related apriori multiplier (187) is
\begin{equation}
m_2(x,\infty) = 1 +\frac{a_1}{a_2}x +\frac{a_2}{a_0}x^2 .
\end{equation}
Starting with $p_2(x)$, let us construct the simplest nested exponentials,
not invoking control functions,
\begin{equation}
f_1^*(x) = a_0\exp\left ( \frac{a_1}{a_0} x\right ) , \qquad
f_2^*(x) = a_0\exp\left ( \frac{a_1}{a_0} x\exp\left (
\frac{a_2}{a_1} x\right )\right ) .
\end{equation}
The sequence $\{ f_k^*(x)\}$ has to satisfy the stability condition (50),
with the aposteriori multipliers (49). In our case, $m_1^*(x)=1$ and
\begin{equation}
m_2^*(x) =\left ( 1 +\frac{a_2}{a_1}x\right )\exp\left (
\frac{a_2}{a_1} x\right ) \frac{f_2^*(x)}{f_1^*(x)} \; .
\end{equation}
Assume now that $a_{n+1}>a_n>0$ and $x>0$. Then $f^*_{n+1}(x)>f^*_n(x)$.
In such a case, multiplier (190) is more than unity. Hence, the procedure
is locally unstable, and we cannot trust to approximants (189). The first
of them, i.e. $f_1^*(x)$, can yet give a reasonable estimate, but the
second, $f_2^*(x)$, is certainly untrustable.

In this way, a quick glance at the apriori multiplier (187) gives us a
feeling of whether the self--similar exponentials would produce good
results. But the final conclusion of whether we did have managed to
construct a convergent sequence of nested exponential approximants is to
be based on the stability analysis of the renormalized multipliers (49).
The relation between the latter and the apriori multipliers (187) is not
direct, as can be seen even from the simplest case resulting in (188) and
(190). In general, it would be more correct to say that there is no direct
relation between these two types of multipliers. Therefore, there is no
necessity of requiring that the apriori multipliers (187) be compulsory
less than unity in their absolute values. It is sufficient to require that
the final multipliers (49) satisfy the stability condition (50).

Another question that may arise is as follows. Suppose, we have met with
the worst case, when we are not able to construct a convergent sequence of
self--similar exponential approximants. How then should we proceed in
order to define an effective limit of a divergent sequence $\{ p_k(x)\}$?
In such a case, there are several possibilities. First of all, since it is
always $m_1^*(x)=1$, as is clear from definition (49), thence $f_1^*(x)$
may serve as an estimate for an effective limit of the sequence 
$\{ p_k(x)\}$.

Second, recall that the nested exponentials are only one of admissible
forms of self--similar approximants, corresponding to a particular case,
when the power in the algebraic transformation (2) is assumed to tend to
infinity. If the latter assumption is waved aside, we return to the
radical form (10). Then, we have to define control functions $s=s_k(x)$
and $\tau=\tau_k(x)$ so that the sequence $\{ p_k^*(x,s_k,\tau_k)\}$ be
convergent. Equations defining these control functions, as always in the
framework of the self--similar approximation theory [7-14], follow from
the requirement for the corresponding approximation cascade to have a
stable fixed point; the stability of a fixed point being in one--to--one
correspondence with the existence of a limit which the sequence of
approximations converges to [10,11,95]. Developing this procedure, we come
to a convergent sequence $\{ f_k^*(x)\}$ of self--similar approximants
$f_k^*(x)=p_k^*(x,s_k(x),\tau_k(x))$ which can give very accurate
approximations for the sought function, but will not have such a nice
structure as that of nested exponentials.

The self--similar exponential approximants, in addition to having a nice
and convenient mathematical structure, evidently illustrate by their form
the idea of self--similarity of approximations [7-9]. Thus, if we
introduce a function
$$ G(x,y) \equiv x e^y $$
and use the notation
$$ x_i\equiv b_ix^{\beta_i} ;\qquad x_0\equiv a_0 x^{\alpha_0} , $$
then the self--similar approximant (18) can be written as
$$ F_k(x) = G(x_0,G(x_1,\ldots G(x_{k-1},x_k))\ldots ) . $$

Finally, one more possibility of treating divergent series, when the
direct construction of nested exponentials does not work, could be either
by resorting to a change of variables or by invoking a transformation of
the given series, so that the following application of the self--similar
approximation theory would result in a convergent sequence of nested
exponentials. What kind of a transformation or a change of variables is
appropriate can again be decided by means of the apriori multipliers (187).

\vspace{5mm}

{\bf Acknowledgment}

\vspace{2mm}

We are grateful to E.P. Yukalova for discussions and advice. We appreciate
financial support from the National Science and Technology Development 
Council of Brazil and from the University of Western Ontario, Canada.

\newpage

{\Large{\bf Appendix}}

The higher--order coefficients in the $\epsilon$--expansion for the 
critical index $\eta$ are [85]
$$
a_4\left( n\right) =-\frac 1{32}\frac{n+2}{(n+8)^6} \times
$$
$$
\times \left[
5n^4+230n^3-1124n^2-17920n-46144+\zeta \left( 3\right) \left( n+8\right)
384\left( 5n+22\right) \right] , 
$$
$$
a_5\left( n\right) =-\frac 1{128}\frac{n+2}{(n+8)^8}\times
$$
$$
\times\left [
\left (13n^6+946n^5+27620n^4+121472n^3-262528n^2-2912768n-5655552\right ) -
\right. 
$$
$$
-\zeta \left( 3\right) \left( n+8\right)
16(n^5+10n^4+1220n^3-1136n^2-68672n-171264)+ 
$$
$$
\left.+\zeta (4)\left( n+8\right) ^31152\left( 5n+22\right) -\zeta (5)\left(
n+8\right) ^25120(2n^2+55n+186)\right ]. 
$$
For the critical index $\nu$ one has
$$
b_3(n)=\frac{(n+2)}{8(n+8)^5}\left[ 3n^3-452n^2-2672n-5312+\zeta
(3)(n+8)96(5n+22)\right] , 
$$
$$
b_4(n)=\frac{(n+2)}{32(n+8)^7}[3n^5+398n^4-12900n^3-81552n^2-219968n-357120 +
$$
$$
+\zeta (3)(n+8)16(3n^4-194n^3+148n^2+9472n+19488) +
$$
$$
+ \zeta (4)(n+8)^3288(5n+22)-\zeta (5)(n+8)^21280(2n^2+55n+186)], 
$$
$$
b_5(n)=\frac{(n+2)}{128(n+8)^9}\times $$
$$
\times\left [
3n^7-1198n^6-27484n^5-1055344n^4-5242112n^3-5256704n^2+6999040n-626688\right. -
$$
$$
-\zeta
(3)(n+8)16(13n^6-310n^5+19004n^4+102400n^3-381536n^2-2792576n-4240640) -
$$
$$
-\zeta ^2(3)(n+8)^21024(2n^4+18n^3+981n^2+6994n+11688) +
$$
$$
+\zeta (4)(n+8)^348(3n^4-194n^3+148n^2+9472n+19488) +
$$
$$
+\zeta (5)(n+8)^2256(155n^4+3026n^3+989n^2-66018n-130608) - 
$$
$$
\left. -\zeta (6)(n+8)^46400(2n^2+55n+186)+\zeta 
(7)(n+8)^356448(14n^2+189n+526)\right ]. $$

The higher--order coefficients of the field--theory expansion [92] for 
the $\beta$--function are
$$
a_4(n)=\frac 1{(n+8)^3}(1.34894276n^2+54.94037698n+199.6404170), 
$$
$$
a_5(n)=-\frac
1{(n+8)^4}(-0.15564589n^3+35.82020378n^2+602.5212305n+1832.206732), 
$$
$$
a_6(n)=\frac1{(n+8)^5}\times $$
$$
\times\left 
(0.05123618n^4+3.23787620n^3+668.5543368n^2+7819.564764n+20770.17697\right ). 
$$
For the critical index $\eta$,
$$
b_4(n)=\frac
1{(n+8)^4}(-0.0042985626n^3+0.6679859202n^2+4.609221057n+6.512109933, 
$$
$$
b_5(n)=-\frac1{(n+8)^5}\left
( 0.0065509222n^4-0.1324510614n^3+1.891139282n^2+15.18809340n+ \right. $$
$$\left.
+ 21.64720643\right ), 
$$
$$
b_6(n)=\frac1{(n+8)^6}\left 
( -0.005548920n^5-0.0203994485n^4+3.054030987n^3+64.07744656n^2+ \right. $$
$$\left. +
300.7208933n+369.7130739\right ). 
$$
For the critical index $\gamma$,
$$
c_3(n)=-\frac 1{(n+8)^3}(0.8795588926n^2+6.485476868n+9.452718166), 
$$
$$
c_4(n)=\frac
1{(n+8)^4}(-0.1283321043n^3+7.966740703n^2+51.84421298n+70.79480631), 
$$
$$
c_5(n)=-\frac1{(n+8)^5}\left (
0.0490966058n^4+4.288152493n^3+108.3618219n^2+537.8136105n+ \right. $$
$$ \left. +
675.6996077\right ), 
$$
$$
c_6(n)=\frac1{(n+8)^6}\left (
-0.0259267945n^5-1.618627843n^4+85.54569746n^3+1538.818235n^2+ \right. $$
$$\left. +
6653.956526n+7862.074086\right ). 
$$

\newpage

\begin{center}
{\bf Table Caption}
\end{center}

Critical indices, for the models with different numbers of components 
$n$, calculated by using self--similar exponential approximants obtained 
from the $\epsilon$--expansion, and compared to the results listed in 
literature. The indices $\nu,\;\eta$, and $\theta$ are calculated 
directly, and other indices are found from the known scaling relations.

\newpage

\begin{table}

\begin{tabular}{|c|c|c|c|c|c|c|c|} \hline
$n$&$\nu$  &$\eta$ &$\gamma$&$\beta$&$\alpha$&$\delta$&$\theta$ \\ \hline
-2 &  1/2  &  0    &  1     & 1/4   &  1/2   &  5         & 0.4    \\ \hline
-1 & 0.545 & 0.019 & 1.08   & 0.278 & 0.365  & 4.888      & 0.431  \\ \hline
   & 0.589 & 0.03  & 1.16   & 0.303 & 0.233  & 4.825      & 0.464  \\
0  & 0.587$\div$0.592 & 0.026$\div$0.034 & 1.157$\div$1.162 
   & 0.302$\div$0.305 & 0.231$\div$0.236& & 0.465$\pm$0.01 \\ 
   & [90 ] & [92 ] & [90 ]  & [92 ] & [92 ]  &            & [91 ] \\ \hline
   & 0.632 & 0.035 & 1.242  & 0.327 & 0.104  & 4.797      & 0.498 \\
 1 & 0.629$\div$0.634 & 0.031$\div$0.038 & 1.237$\div$1.244
   & 0.324$\div$0.327 & 0.107$\div$0.110& & 0.5$\pm$0.02\\
   & [90 ] & [92 ] & [90 ]  & [92 ] & [92 ]  &            &  [91 ]\\ \hline
   & 0.671 & 0.036 & 1.318  & 0.348 &$-$0.013 & 4.792      & 0.531 \\
 2 & 0.662$\div$0.677 & 0.032$\div$0.039 & 1.308$\div$1.327
   & 0.346$\div$0.348 &$-$0.007$\div-$0.1& & 0.52$\pm$0.02\\
   & [90 ] & [92 ] & [90 ]  & [92 ] & [92 ]  &            & [91 ]\\ \hline
   & 0.708 & 0.037 & 1.39   & 0.367 &$-$0.124 & 4.786      & 0.562 \\
 3 & 0.704$\div$0.72 & 0.031$\div$0.038 & 1.385$\div$1.406
   &0.362$\div$0.366 &$-$0.115$\div$$-$0.117 & & 0.55$\pm$0.015\\
   & [90 ] & [92 ] & [90 ]  & [92 ] & [92 ]  &            & [91 ]\\ \hline
   & 0.741 & 0.036 & 1.455  & 0.384 &$-$0.223 & 4.792      & 0.592 \\
 4 &0.738$\div$0.755 & 0.036 & 1.449$\div$1.483 & 0.382 &$-$0.213 & & \\
   & [90 ] & [92 ] & [90 ]  & [92 ] & [92 ]  &            & \\ \hline
   & 0.797 & 0.033 & 1.568  & 0.412 &$-$0.391 & 4.808     & 0.645 \\
 6 & 0.79$\div$0.818 & 0.031 & 1.556$\div$1.608 & 0.407 &$-$0.37 & & \\
   & [90 ] & [92 ] & [ 90]  & [92 ] & [92 ]  &            & \\ \hline
   & 0.84  & 0.029 & 1.656  & 0.432 &$-$0.52  & 4.831      & 0.688 \\
 8 & 0.83$\div$0.856&0.027 & 1.637$\div$1.687 & 0.426 &$-$0.489 & & \\
   & [90 ] & [92 ] & [90 ]  & [92 ] & [92 ]  &            &  \\ \hline
   &0.872  & 0.026 & 1.721  & 0.447 &$-$0.616 & 4.848      & 0.723 \\
10 & 0.85$\div$0.884 & 0.024 & 1.697$\div$1.744 & 0.440 &$-$0.576 & & \\
   & [90 ] & [92 ] & [90 ]  & [92 ] & [92 ]  &            &  \\ \hline
   & 0.896 & 0.023 & 1.771  & 0.458 &$-$0.688 & 4.865      & 0.751 \\
12 & 0.881$\div$0.902 & 0.021 & 1.741$\div$1.783 & 0.450 &$-$0.643 & &  \\
   & [90 ] & [92 ] & [90 ]  & [92 ] & [92 ]  &            & \\ \hline
$\infty$& 1 & 0    & 2      & 1/2   & $-$1   & 5          & 1 \\ \hline
\end{tabular}

\end{table}

\end{document}